\documentclass[authoryear,12pt,letter,3p]{jowarticle}
\usepackage{graphicx}
\usepackage{amsmath}
\usepackage{amssymb}
\usepackage{setspace}
\usepackage[all]{xy}
\makeatletter
\renewcommand\section{\@startsection{section}{1}{\z@}{-3.25ex plus -1ex minus -.2ex}{1.5ex plus .2ex}{\normalsize\bf}}
\renewcommand\subsection{\@startsection{subsection}{2}{\z@}{-3.25ex plus -1ex minus -.2ex}{1.5ex plus .2ex}{\normalsize\bf}}
\renewcommand\subsubsection{\@startsection{subsubsection}{3}{\z@}{-3.25ex plus -1ex minus -.2ex}{1.5ex plus .2ex}{\normalsize\bf}}
\makeatother

\newtheorem{thm}{Theorem}

\newtheorem{prop}[thm]{Proposition}

\newtheorem{con}{Condition}

\begin{document}
\begin{frontmatter}
\title{Against Dogma: On Superluminal Propagation in Classical Electromagnetism}
\author{James Owen Weatherall}\ead{weatherj@uci.edu}
\address{Department of Logic and Philosophy of Science\\ University of California, Irvine, CA 92697}
\begin{abstract}It is deeply entrenched dogma that relativity theory prohibits superluminal propagation.  It is also experimentally well-established that under some circumstances, classical electromagnetic fields propagate through a dielectric medium with superluminal group velocities and superluminal phase velocities.  But it is usually claimed that these superluminal velocities do not violate the relativistic prohibition.  Here I analyze electromagnetic fields in a dielectric medium within a framework for understanding superluminal propagation recently developed by \citet{Geroch1,Geroch2} and elaborated by \citet{Earman}.  I will argue that for some parameter values, electromagnetic fields \emph{do} propagate superluminally in the Geroch-Earman sense.\end{abstract}
\end{frontmatter}

\doublespacing

\section{Introduction}

Few dogmas in modern physics are as well-entrenched as the one stating that relativity theory prohibits superluminal propagation.  And yet, despite its crucial importance to many physical arguments---foundational and otherwise---it is not fully clear what the status of this would-be prohibition is within relativity theory.  Is it physical fields, such as electromagnetic fields or Klein-Gordon fields, that cannot propagate superluminally?  Or is it energy-momentum?  Or is it some variety of superluminal \emph{signaling} that is prohibited?  If the latter, then is there some unambiguous physical criterion for what constitutes a signal, or does signaling depend essentially on the possible transmission of information---perhaps between intentional beings?  And whatever the details, is the prohibition on superluminal propagation supposed to be a \emph{consequence} of relativity?  Or is it a brute assumption, independent of the rest of the theory?

These questions are not idle quibbling about definitions.  The idea that relativity theory, in some sense or another, prohibits superluminal propagation directly influences physicists' theoretical understanding of physical processes and their interpretation of experimental results.  The prohibition also plays a central role in attempts to reconcile quantum physics with relativity.  Moreover, there are concrete cases where the ambiguity concerning precisely what it is that relativity is meant to prohibit has led to confusion in the physics literature.  For instance, in the context of experiments concerning light pulses in dielectric media, which I will discuss in more detail below, various apparently superluminal effects have been observed.\footnote{Here I limit attention to cases where the electromagnetic field is treated classically.  Examples of purportedly superluminal phenomena multiply if one considers quantum electrodynamics.  See \citet{Butterfield} for a discussion of these examples aimed at philosophers.}  In such cases, it is ubiquitous practice to provide some argument for why the observed superluminal phenomena do not constitute superluminal propagation of a sort that would conflict with relativity.  But these arguments have a decidedly \emph{ad hoc} flavor and relatively little attention is paid, at least in this literature, to the more principled questions of what \emph{would} constitute superluminal propagation of the troubling sort and how, in these particular cases, relativity manages to forbid it.  At the very least, although relativity is often mentioned, a satisfactory relativistic treatment of the systems in question is rarely, if ever, on offer.

This is not to say that the more principled question is never taken up.  In recent work, \citet{Geroch2} and \citet{Earman} have articulated a precise and general account of what it would mean for a physical system to propagate superluminally in relativity theory.\footnote{\citet{Weinstein} may be seen as a sympathetic precursor to the view recently defended by Geroch and Earman.  The principal difference, if one exists at all, concerns the role of ``causal cones'' (see section \ref{Geroch}, below) in the criterion of (maximal) field propagation velocity.}  More strikingly, Geroch, at least, argues that such fields should be understood as compatible with relativity theory, and both Earman and Geroch present examples of fields that are, in some sense, ``relativistic,'' and yet which exhibit superluminal propagation according to their criterion.\footnote{So as not to besmirch their good names, let me emphasize that neither Geroch nor Earman suggests that there are physical systems that \emph{do} propagate superluminally---and indeed, Earman takes the upshot of the discussion to be a more precise characterization of what we intend relativity to prohibit, as a guide to building a prohibition on superluminal propagation into relativistic quantum field theory.}

My goal in the present paper is simply to bring these two literatures together.  Along the way, I will defend three theses.  First, I will argue that the sense of superluminal propagation developed by Geroch and elaborated by Earman can be understood as making precise a notion of propagation already present in the literature on electromagnetic fields in a dielectric---namely, superluminal values of the so-called Sommerfeld-Brillouin ``wavefront velocity'' associated with a field.\footnote{\label{signal} This quantity is often called the ``signal velocity'' in the literature.  It is interesting to note, however, that Sommerfeld himself distinguishes the wavefront velocity he defines from the signal velocity (i.e., the group velocity) that Brillouin discusses \citep[p. 19]{Brillouin}.  I will follow Sommerfeld and call this quantity the ``wavefront velocity''.}  Second, I will argue that in at least one highly idealized case, on a fully relativistic treatment, electromagnetic fields governed by the equations of motion for an electromagnetic field in a dielectric \emph{do} propagate superluminally in the Geroch-Earman sense. Finally, I will argue that an oft-cited argument due to \citet{Sommerfeld} intended to show that superluminal wavefront velocities are impossible has nothing to do with relativity \emph{per se}, and instead gains what force it has from an assumption concerning the nature of the interaction between electromagnetic fields and matter motivated by the atomic theory of matter.

Let me also emphasize what I am \emph{not} arguing.  I do not mean to argue that there are physical systems that, under realistic conditions, \emph{do} exhibit superluminal propagation in the Geroch-Earman sense.  Nor do I mean to argue that it is possible to engineer a dielectric medium through which one could send a signal superluminally, let alone that such media have already been produced.  In this way, the title of the paper may be misleading, as I do not mean to argue that the dogma noted above is false.  But I do hope to show that we do not understand the relationship between relativity theory and superluminal propagation as well as we might think, even in cases of manifest physical interest (insofar as we have built components of optical systems that exhibit some of the relevant properties).  And in this sense, the dogma that relativity theory simply forbids superluminal propagation is unhelpful.  On the one hand, it discourages study of \emph{how} relativity theory does and does not accommodate superluminal propagation.  And perhaps worse, it may blind us to systems that \emph{do} exhibit superluminal propagation in physically significant and potentially fruitful ways.

The rest of the paper will proceed as follows.  I will begin with some preliminaries regarding Maxwell's equations, to establish notation and conventions, and to provide a translation manual between different ways of presenting Maxwell's theory.  Next I will reconstruct several standard arguments concerning superluminal propagation of electromagnetic fields in a dielectric. I will then present Geroch's framework for treating the propagation of fields and argue that his approach provides a natural way of precisely recovering the Sommerfeld-Brillouin notion of wavefront velocity.  Using this framework, I will analyze the standard relativistic field equations for an electromagnetic field in a dielectric medium and show that for certain parameter values, these fields will exhibit superluminal propagation in the Geroch-Earman sense---i.e., they will have superluminal wavefront velocities.  Finally, I will return to Sommerfeld's no-go argument for superluminal wavefront velocities and discuss how the example I present runs afoul of his assumptions.  The upshot will be that insofar as Sommerfeld's argument succeeds, relativity theory plays no apparent role.  I will conclude by stating, and to some extent responding to, a number of objections to the analysis I give and suggesting avenues for future work.

\section{Preliminaries}\label{prelim}

In what follows, we work in Minkowski spacetime, $(M,\eta_{ab})$, in units in which the speed of light, $c$, equals 1 (though, for emphasis, we will sometimes continue to refer to $c$ as the speed of light).\footnote{\label{AIN} Minkowski spacetime $(M,\eta_{ab})$ is a relativistic spacetime where $M$ is $\mathbb{R}^4$ and $\eta_{ab}$ is flat and geodesically complete.  Throughout we use the ``abstract index'' notation developed by \citet{Penrose+Rindler} and used by \citet{Wald} and \citet{MalamentGR}.  We adopt the convention that the Minkowski metric has signature $(1,3)$, so that timelike vectors have positive inner product with themselves.}  We assume that Minkowski spacetime is endowed with a fixed temporal orientation and a fixed orientation, with associated volume element $\epsilon_{abcd}$.  In this context, Maxwell's equations for electromagnetic fields in a vacuum may be written in a compact form as:
\begin{subequations}\label{Maxwell1}
\begin{align}
\nabla_aF^{a}{}_{b}&=J_b\label{inhomo}\\
\nabla_{[a}F_{bc]}&=\mathbf{0},\label{closed}
\end{align}
\end{subequations}
where $\nabla$ is the Minkowski spacetime derivative operator, $F_{ab}$ is the electromagnetic field tensor, and $J^a$ is the charge-current density.  Indices are raised and lowered with $\eta_{ab}$.

The electromagnetic field $F_{ab}$ can be taken to encode the electric and magnetic fields as determined by any observer, as follows.  Given an observer with 4-velocity $\xi^a$ at a point $p$ in Minkowski spacetime, the electric field determined by that observer is given by $E^a=F^a{}_b\xi^b$ and the magnetic field is given by $B^a=\frac{1}{2}\epsilon^{abcd}\xi_b F_{cd}$.  Similarly, $\sigma=J_a\xi^a$ is the charge density as determined by that observer, whereas $j^a=J^a - (J^n\xi_n)\xi^a$ is the 3-current density determined by that observer.

It will be convenient to be able to move back and forth between this manifestly relativistic form of Maxwell's equations and a more traditional formulation, which is more common in the literature on the propagation of electromagnetic waves.\footnote{For further details on the relationship between these formulations, see \citet{MalamentGR}.  When I say ``relativistic'' in this setting, I mean independent of a choice of observer or coordinate system.}  To do so, we will fix, once and for all, a constant future-directed unit timelike vector field $\xi^a$ on Minkowski spacetime, representing, say, the 4-velocities of a family of co-moving inertial observers.  Unless otherwise stated, the electric and magnetic fields, $E^a$ and $B^a$, and the charge and current 3-vector densities, $\sigma$ and $j^a$, will always be assumed to be determined relative to this family of observers.  The electromagnetic field tensor $F_{ab}$ can be reconstructed in terms of these fields as:
\begin{equation}
F_{ab}=2E_{[a}\xi_{b]} + \epsilon_{abnm}\xi^n B^m.
\end{equation}

Eqs. \eqref{Maxwell1} then can be re-written as
\begin{subequations}\label{Maxwell2}
\begin{align}
\nabla_{[a}F_{bc]}&=0 & &\Longleftrightarrow& & \begin{cases} & \partial_b B^b = 0 \\
& \epsilon^{abc} \partial_b E_c = -\xi^b\nabla_b B^a  \end{cases}\\
\nabla_a F^{a}{}_{b}&=J_b & &\Longleftrightarrow& & \begin{cases}{} & \partial_b E^b = \sigma \\
&  \epsilon^{abc} \partial_b B_c = \xi^b\nabla_b E^a + j^a\end{cases}
\end{align}
\end{subequations}
where $\epsilon_{abc}=\epsilon_{abcn}\xi^n$ is the induced volume element on three dimensional hypersurfaces orthogonal to $\xi^a$ and $\partial_a = h^n{}_a \nabla_n$, where $h^a{}_b=\delta^a{}_b - \xi^a\xi_b$ is the projection onto these hypersurfaces, is the induced derivative operator on the hypersurfaces determined by $\nabla$.

In yet another notation, Eqs. \eqref{Maxwell2} are just
\begin{subequations}\label{Maxwell3}
\begin{align}
&\partial_b B^b = 0& &\Longleftrightarrow& &\vec{\nabla}\cdot\mathbf{B} = 0&\\
&\epsilon^{abc} \partial_b E_c = -\xi^b\nabla_b B^a& &\Longleftrightarrow& &\vec{\nabla}\times\mathbf{E} = -\frac{\partial \mathbf{B}}{\partial t}&\\
&\partial_b E^b = \sigma& &\Longleftrightarrow& &\vec{\nabla}\cdot\mathbf{E} = \sigma &\\
&\epsilon^{abc} \partial_b B_c = \xi^b\nabla_b E^a + j^a& &\Longleftrightarrow& &\vec{\nabla}\times\mathbf{B} = \frac{\partial \mathbf{E}}{\partial t} + \mathbf{j}&
\end{align}
\end{subequations}
where now $\vec{\nabla}$ is the standard gradient operator on the $\mathbb{R}^3$ hypersurfaces orthogonal to $\xi^a$, and the time derivative is taken relative to the global time coordinate determined by $\xi^a$.

Several standard arguments are available to show that in the absence of any sources---i.e., for $J_a=\mathbf{0}$, which we will assume from now on---Maxwell's equations may be understood as a wave equation for waves propagating along null geodesics---i.e., curves whose 3-velocity as determined by any observer would be $c$.  For instance, following \citet{Wald}, one may observe that since Minkowski spacetime is contractible, Eq. \eqref{closed} implies that there exists a globally defined 1-form $A_a$ such that $F_{ab}=2\nabla_{[a}A_{b]}$.  This relation determines $A_a$ only up to the derivative of a smooth scalar field, since for any such field $\psi$, $\nabla_{[a}\nabla_{b]}\psi=\mathbf{0}$.  It follows that one may always choose the field $A_a$ such that $\nabla_a A^a=0$.  This choice is known as the ``Lorenz gauge''.  Writing Eq. \eqref{inhomo} in terms of this field $A_a$ yields
\begin{equation}\label{wave1}
\mathbf{0}=\nabla_nF^{na}=\nabla_n\nabla^nA^a-\nabla_n\nabla^a A^n = \nabla_n\nabla^n A^a,
\end{equation}
where we have used the fact that Minkowski spacetime is flat to commute the derivative operators in deriving the last equality.

Eq. \eqref{wave1} is a wave equation for the vector potential.  But the physical significance of this equation may not be immediately clear, since the vector potential is not usually taken to have direct physical significance.  Note, however, that Eq. \eqref{wave1} implies that,
\begin{equation}\label{wave2}
\nabla_n\nabla^n F_{ab}=\mathbf{0},
\end{equation}
which in turn is equivalent to
\begin{subequations}\label{wave3}
\begin{align}
\nabla_n\nabla^n E^a &= \mathbf{0}\label{efieldwave}\\
\nabla_n\nabla^n B^a &= \mathbf{0}
\end{align}
\end{subequations}
These last two expressions may also be written as
\begin{subequations}\label{wave4}
\begin{align}
\square\mathbf{E}=(c^2\frac{\partial^2}{\partial t^2} - \vec{\nabla}^2)\mathbf{E}&=\mathbf{0}\\
\square\mathbf{B}=(c^2\frac{\partial^2}{\partial t^2} - \vec{\nabla}^2)\mathbf{B}&=\mathbf{0}
\end{align}
\end{subequations}
where we have introduced $c$ for clarity.  (Eqs. \eqref{wave4} can also be derived by applying standard identities from vector calculus to Eqs. \eqref{Maxwell3}.)

Eqs. \eqref{wave1}-\eqref{wave4} may all be understood as systematically related wave equations.  They all admit wave-like solutions of the same characteristic form.  For instance, taking Eq. \eqref{efieldwave} as an example,\footnote{The other wave equations described have solutions of the same form, differing only in the details of the amplitude vector/tensor.} we have solutions of the form
\begin{equation}\label{waveSol1}
E^a=C^a e^{i S}
\end{equation}
where $S$, a scalar field, is the \emph{phase} of the wave and $C^a$ is the (constant) amplitude of the wave.\footnote{In Eq. \eqref{waveSol1} and throughout the paper, we take for granted that electromagnetic fields are represented by the \emph{real part} of any complex quantities defined.  In all discussions of Fourier analysis, for instance, we implicitly restrict attention to only the real parts of integrals over complex exponentials.}  A straightforward calculation (see \citet[pp. 65-6]{Wald}) shows that Eq. \eqref{wave1} implies that $k_a\equiv\nabla_a S$, the normal vector to surfaces of constant phase, is null; thus $k^a$ is also tangent to the surfaces of constant phase.\footnote{To clarify: since null vectors have zero inner product with themselves, they count as orthogonal to themselves.  Thus if the covector $k_a$ normal to a $3-$surface is null, then the vector $k^a=\eta^{ab}k_b$ will be tangent to the $3-$surface, since its action on $k_a$ vanishes.}  Moreover, $k^a$ can be shown to be geodesic, i.e., $k^n\nabla_n k^a=\mathbf{0}$.  This provides a sense in which solutions of the form of Eq. \eqref{waveSol1} propagate along null geodesics, insofar as points of constant phase in the waveform may be said to propagate along such curves.  And thus, since null curves have velocity $c$ relative to any observer, we have a sense in which these solutions propagate at the speed of light.

The field $k^a$ has a natural interpretation as the 4-momentum density associated with an electromagnetic wave.  As with any 4-momentum density, we can decompose $k^a$ relative to $\xi^a$.  We identify $\omega=k^a\xi_a$, the energy density relative to $\xi^a$, as the frequency of the wave, and $\hat{k}^a=k^a-(k^n\xi_n)\xi^a$, the 3-momentum, as the wave vector.  (We will denote this latter quantity by $\mathbf{k}$ when we want to emphasize that it is a 3-vector.)  Note that since $k^a$ is null, $k^ak_a = \omega^2 - c^2\mathbf{k}\cdot\mathbf{k}=0$, and thus $c=\omega/|\mathbf{k}|$, which is the \emph{dispersion relation} relating the frequency, 3-momentum, and speed of a wave-like solution of Maxwell's equations.

The most important special case of wave-like solutions is that of the monochromatic plane waves.  For any (fixed) point $p$, these may be written as
\[
E^a=C^a e^{-i k^n\overset{p}{\chi}_n}=C^a e^{i (\hat{\mathbf{k}}\cdot\mathbf{x}-\omega t)}
\]
where $\overset{p}{\chi}{}^a$ is the position vector field centered at $p$,\footnote{The position vector field is the unique field on Minkowski spacetime such that (1) $\nabla_a\overset{p}{\chi}{}^b=\delta_a{}^b$ and (2) $(\overset{p}{\chi}{}^a)_{|p}=\mathbf{0}$.  One can think of it as the field that assigns to each point $q$ the (parallel transport of) the vector at $p$ connecting $p$ to $q$.  In this way, it records data about a global coordinate system centered at $p$ in a coordinate-independent manner.  See \citet[p. 66]{MalamentGR} for more on the position vector field.} and where $t$ and $\mathbf{x}$ are (standard global) coordinates with origin at $p$.   A basic result of Fourier analysis is that any solution of Maxwell's equations with sufficiently nice properties at infinity may be represented as an integral over plane wave solutions, with the general form
\begin{equation}\label{genSol}
E^a(\mathbf{x},t) = \frac{1}{\sqrt{2\pi}}\int_{\mathbb{R}^3} d\hat{\mathbf{k}} C^a(\hat{\mathbf{k}}) e^{i(\hat{\mathbf{k}}\cdot\mathbf{x}-\omega(\hat{\mathbf{k}}) t)},
\end{equation}
where
\begin{equation}\label{genCoef}
C^a(\hat{\mathbf{k}})=\frac{1}{\sqrt{2\pi}}\int_{\mathbb{R}^3}d\mathbf{x} E^a(\mathbf{x},0)e^{-i\hat{\mathbf{k}}\cdot\mathbf{x}}.
\end{equation}
Thus, modulo behavior at infinity, any solution, wave-like or not, may be understood to consist of a linear superposition of waves that all propagate at velocity $c$.  This provides an even more robust sense in which one might say that electromagnetic fields in a vacuum propagate at $c$.\footnote{One might still worry that these arguments are not quite as robust as one would like.  For instance, it is not clear how the constraints on the behavior of fields at infinity required for Fourier analysis are physically motivated, particularly in the source-free case.  (If one assumes all fields are generated by localized sources, one might argue that fields should vanish at infinite distance from the sources.)  Fortunately, other arguments, not subject to such limitations, also exist to establish that electromagnetic fields should be said to propagate at the speed of light in a vacuum.  We will describe one more in detail in section \ref{Geroch}, and then point to yet another (related) argument in section \ref{Sommerfeld}.  But since our purpose here is merely to establish notation and set up some basic facts that will be necessary for what follows, we will break off the current discussion here, and proceed to the main arguments of the paper.}

\section{A Dielectric Dialectic}\label{Milonni}

We have now seen a sense in which a broad class of electromagnetic fields may be said to propagate at $c$.  But all of the considerations raised so far apply only in a vacuum; in other contexts, these arguments do not apply, and indeed, it is fruitful in some cases to define different velocities for light.\footnote{The \emph{locus classicus} for discussions of wave propagation in dielectric media is \citet{Brillouin}.  See also \citet{Born+Wolf}, \citet{Oughstun+Sherman}, and, especially, \citet{Milonni}, which discusses the experimental literature up to the book's publication and also provides a detailed discussion of the various senses in which light may and may not propagate superluminally.}  For instance, light waves passing through a transparent medium, such as water or glass, will generally \emph{refract}, or bend.  This behavior may be understood as a change in the relationship between the frequency $\omega$ and the 3-momentum $\mathbf{k}$ of a wave propagating through the medium, so that for a given monochromatic plane wave, one finds
\[
\frac{\omega}{|\mathbf{k}|}\mapsto n\left(\frac{\omega}{|\mathbf{k}|}\right),
\]
where $n$ is known as the \emph{index of refraction}.\footnote{In general, the index of refraction will be a complex function of the frequency of a wave incident on the medium.  The imaginary part of the index encodes information about the absorptive properties of the medium; the real part determines the velocities we discuss here.  Whenever ``$n$'' appears in an expression in what follows, it should be assumed that it is the real part of the index that is under discussion; alternatively, one might assume we are only working in transparent frequency bands, where the imaginary part of the index (approximately) vanishes.}

Since the fact that $k^a$ is null in a vacuum implies $\omega/|\mathbf{k}|=c$, a refractive medium may be interpreted as changing the velocity of light, such that now we have $\omega/|\mathbf{k}|=c/n\equiv_{\text{def}} v_p$, where $v_p$ is the \emph{phase velocity} of the plane wave in the medium.\footnote{This value may be thought of as a ``velocity'' for waves propagating in one spatial dimension; otherwise, it should really be thought of as a phase \emph{speed}, since it is not a vector.  However, we will adopt the standard usage and call this the phase velocity.\label{speed1}}  The phase velocity is a measure of the velocity of a point of constant phase in a wave-form, for a single frequency (monochromatic) wave.  In this sense, the notion of speed given by the phase velocity is precisely the one captured by the discussion in the previous section.  We may think of the index of refraction of the vacuum, then, as 1.  For familiar media, such as water or glass, at optical frequencies---i.e., plane wave frequencies we associate with visible light---the index of refraction is greater than 1.  This immediately implies that the phase velocity of light at these frequencies passing through such media slows relative to $c$.  But it is also experimentally well-established that in some media---including, for instance, glass at x-ray frequencies, as well as various engineered media---the index of refraction may fall below 1.  When this occurs, the phase velocity becomes superluminal.

Such media are not new, and it has long been known that electromagnetic fields may have superluminal phase velocities.  But this is usually not treated as a violation of the relativistic prohibition on superluminal propagation.  Several reasons are given in the literature.  For instance, \citet{Milonni} writes, ``[phase velocity] is associated with monochromatic waves and, therefore, can be greater than $c$ without violating special relativity'' (p. 58).  The idea here seems to be that phase velocity depends essentially on an idealization---that of single-frequency (monochromatic) plane waves---and is therefore unphysical.   A second argument is that a monochromatic plane wave cannot carry ``information''.  The reason is that a monochromatic plane wave is, by definition, of infinite extent, in the sense that it is non-zero on an unbounded region of spacetime.  Moreover, the waveform is completely determined by the values of the wave on small regions of spacetime.  Thus, were a monochromatic plane wave present in a region of spacetime, the values of the field at all other regions would be fixed, meaning that one could not use variations in the field to transmit information (while preserving the assumption that the field is a monochromatic plane wave).  For these reasons, although one can consistently assign a velocity to monochromatic plane waves, their defining properties undermine the idea that such waves exhibit ``propagation'' in the salient sense.\footnote{There is good reason to resist both of these arguments.  The second, for instance, amounts to the claim that because we cannot use plane waves to signal, they do not propagate, which seems to be a non-sequitur.  Nonetheless, such arguments are present in the literature and form part of the motivation for looking to other notions of wave velocity, so I report them here.}

Insofar as these considerations are convincing, they suggest that one should look elsewhere if one is interested in characterizing a sense of ``propagation velocity'' for (realistic) electromagnetic fields.  And indeed, electrical engineers working on, say, radio communication \emph{do} look elsewhere.  As we saw in the previous section, a broad class of solutions to Maxwell's equations may be understood as superpositions of monochromatic plane waves.  Insofar as such fields are used to send signals---or rather, insofar as they ``propagate'' in a physically interesting sense---it turns out to be variations in these superpositions that matter, not the propagation of the individual plane waves.  In many cases, we might think of these superpositions as forming ``bumps'' or ``packets'' or ``peculiarities'' (Lord Rayleigh's term) in a wave form.\footnote{See \citet{Rayleigh}.  It seems to me that ``pecularities'' biases the discussion less than the other terms, and so I will use it in what follows; I recognize that this is idiosyncratic by contemporary standards!}  Thus---the standard argument goes---we are not interested in phase velocity; rather, we are interested in the \emph{group velocity},\footnote{As with phase velocity, a better expression for the quantity I will presently define might be ``group speed,'' since it is not vectorial except in one spatial dimension.  Recall footnote \ref{speed1}.} which is a measure of the propagation speed of these peculiarities---at least in the special case where the peculiarity may be conceived as a superposition of a group of plane waves whose 3-momenta are proportional, but with frequency and 3-momentum length varying within a ``small'' range.  The group velocity is distinct from the phase velocity whenever \emph{dispersion} is present, i.e., whenever the index of refraction depends on the frequency of a monochromatic plane wave in a medium.

The group velocity is typically defined as:
\begin{equation}
v_g=\frac{d\omega}{d|\mathbf{k}|}.
\end{equation}
Strictly speaking, this expression does not capture what we want: in general, this derivative will be a function of the 3-momentum, whereas we want a ``group velocity'' to assign a single value to a ``group'' of plane waves with different frequencies/3-momenta.  And so what one really wants to do is evaluate this function at some ``central'' value of the 3-momentum, usually determined by physical considerations related to the problem at hand, to get a determinate value for the group velocity.\footnote{In the language of signal processing in engineering, one typically evaluates the expression at the frequency of the ``carrier wave''.}  

Using the fact that $|\mathbf{k}|=\frac{n\omega}{c}$, we can rewrite the group velocity as a function of the derivative of the index of refraction, as:\footnote{To see this, differentiate both sides with respect to $|\mathbf{k}|$ to find $c=\frac{dn}{d|\mathbf{k}|}\omega+\frac{d\omega}{d|\mathbf{k}|}n=\frac{dn}{d\omega}\frac{d\omega}{d|\mathbf{k}|}\omega+\frac{d\omega}{d|\mathbf{k}|}n=\left(\frac{dn}{d\omega}\omega+n\right)v_g$.}
\begin{equation}\label{groupv2}
v_g=\frac{c}{n+\omega\frac{dn}{d\omega}},
\end{equation}
where once again the right-hand side is evaluated at some ``central'' frequency.  Eq. \eqref{groupv2} clarifies the relationship between group velocity and phase velocity.  In particular, it shows that when $\frac{dn}{dw}=0$, i.e., when there is no dispersion, group velocity and phase velocity coincide.  Meanwhile, for frequencies of ``normal dispersion'', which corresponds to $\frac{dn}{d\omega}>0$, the group velocity will always be less than the phase velocity, which means that when the phase velocity becomes superluminal, the group velocity may still be subluminal.\footnote{Though one might still worry: after all, there is no guarantee that the group velocity will be subluminal if the phase velocity is superluminal---just that that group velocity will less than the phase velocity.}  And thus, the argument goes, relativity is saved from superluminal phase velocities by corresponding subluminal group velocities.

I should emphasize that the group velocity is an essentially approximate notion.  It is a useful way of capturing the velocity of a peculiarity only when the peculiarity propagates through a medium without significant change of shape or amplitude.  This means that group velocity is only salient when the medium is transparent to fields in the relevant frequency band and the index of refraction depends approximately linearly on the frequency of the wave, so that $\frac{dn}{d\omega}$ may be treated as a constant in Eq. \eqref{groupv2}.  Moreover, it is not clear that the group velocity, which depends on a preferred basis in Minkowski spacetime determined by the 4-velocity of the medium and the direction of propagation of the wave, could be extended to an invariant 4-velocity in a fruitful way.  (This is not a worry for the phase velocity, which may be defined in terms of the tangents to the surfaces of constant phase for certain electromagnetic fields in a medium; in contrast, the group velocity depends on facts about the interference of different monochromatic plane waves and the overall shape of a peculiarity, determinations of which will in general vary from observer to observer.)  For these reasons, one might be cautious about assigning much foundational significance to the group velocity.

But even if one does take the group velocity to resolve worries about superluminal propagation of electromagnetic fields in cases of normal dispersion, it is a temporary victory at best, since in the presence of so-called ``anomalous dispersion,'' where $\frac{dn}{d\omega}<0$, the group velocity exceeds the phase velocity.  And there exist media in which the dispersion is anomalous and of sufficient magnitude that the group velocity becomes superluminal.  Indeed, evidence of superluminal group velocities was observed experimentally as early as 1970, by \citet{Faxvog+etal}; more recently, superluminal group velocities were observed in the absence of any significant change in the shape or amplitude of the peculiarity by \citet{Wang+etal}.  These latter experiments in particular, which have been reproduced in various forms by several groups, are usually taken to establish that superluminal group velocities are possible in physical media.\footnote{For a detailed overview of the state of the experimental literature, see \citet[Ch. 2]{Milonni} and \citet{WeatherallDiss}.}

Recall that one reason that superluminal phase velocities were deemed untroubling for relativity theory was that monochromatic plane waves cannot be used to transmit information.  But pecularities in a wave form surely \emph{can} be used to transmit information (think of AM radio!)---which is precisely what motivated the move to group velocity in the first place.  Indeed, in the early history of relativity theory, the possibility of superluminal group velocities was a matter of considerable concern.  As Brillouin put it in the preface to his treatise on wave propagation,
\begin{quote}\singlespacing
...the theory of relativity ...  states that no velocity can be higher than $c$, the velocity of light in vacuum.  Group velocity, as originally defined, became larger than $c$ or even negative within an absorption band.  Such a contradiction had to be resolved and was extensively discussed in many meetings about 1910. \citep[p. vii]{Brillouin}
\end{quote}
Today, however, superluminal group velocities are widely viewed as unproblematic from the perspective of relativity theory, largely on the basis of an argument due to \citet{Sommerfeld}, which was offered as a response to these early worries.

Sommerfeld's position was that one should really consider yet another notion of the velocity of a wave,\footnote{Or rather, again, speed.} namely the velocity of a \emph{wavefront}:
\begin{quote}\singlespacing
In order to say something about propagation, we must ... have a limited wave motion: nothing until a certain moment in time, then, for instance, a series of regular sine waves, which stop after a certain time or which continue indefinitely.  Such a wave motion will be called a \emph{signal}. Here, one can speak of a propagation of the front of the wave (wavefront velocity).... \citep[p. 18]{Brillouin}
\end{quote}
It is this wavefront velocity---the velocity, as Sommerfeld goes on to argue, of a \emph{jump discontinuity} in a solution to Maxwell's equations---that Sommerfeld claims is the salient one, at least as far as the relativistic prohibition is concerned.\footnote{See fn. \ref{signal}.  It has subsequently become common practice to call the wavefront velocity the ``signal velocity,'' but I avoid that usage here.  Note, too, that in the next section, we will generalize to any discontinuity in a field or its derivatives.}  The idea is that, whatever else is the case, a peculiarity in a wave---and thus, any information encoded therein---cannot reach a detector before the wave itself reaches the detector.  And thus, there is a sense in which the wavefront velocity may be understood as the upper bound on the velocity of any signal encoded in an electromagnetic field.  Sommerfeld then presented an argument that the wavefront velocity is always equal to $c$, irrespective of the optical properties of the medium in which the wave is propagating.  (I will discuss this argument in detail in section \ref{Sommerfeld}; for now I simply record that it is widely taken to have settled the matter.)

How are we to interpret superluminal group velocities, then?  On the final accounting, it seems systems exhibiting superluminal group velocities are best conceived in terms of dynamical reshaping of an electromagnetic field, in such a way that some particular feature of the wave appears to propagate at a superluminal velocity.  For instance, if one considers a pulse propagating through a medium, then the leading tail might be amplified, while the pulse itself is damped, in such a way that the leading tail comes to have qualitative features we would have associated with the pulse, while the original pulse disappears.  But this sort of reshaping can occur only when the wave is, in a sense, already present in a medium.  And so, the argument concludes, superluminal group and phase velocities---and electromagnetic radiation more generally---present no problems for the relativistic prohibition on superluminal propagation, since only wavefront velocity matters, and this velocity is always precisely $c$.

\section{Geroch and Earman on Superluminal Propagation}\label{Geroch}

In the previous section, I have attempted to present various senses in which electromagnetic fields do and do not exhibit superluminal propagation, in the terms usually discussed in the literature on propagation in a dielectric medium.  In this section I will turn to a quite different, and in some respects more principled, treatment of field propagation that applies to essentially \emph{any} physically salient system of fields on a manifold.  The analysis I will presently describe, due to \citet{Geroch1}, is manifestly geometrical and relativistic, at least in the sense of being coordinate independent.\footnote{The details of the formalism I will present here are developed by Geroch, but there is a sense in which he merely re-packages the already well-established theory of hyperbolic systems of differential equations.  (See \citet{John}, \citet{Lax}, or \citet{Evans} for more traditional presentations.)  For present purposes, the principal virtue of his approach is that it avoids worries that the notion of field propagation he describes is coordinate or frame dependent.  Such worries at least apparently arise on standard presentations of hyperbolic systems theory.}  I will begin by translating Maxwell's equations into this formalism.  Then I will present some general definitions and propositions that allow one to define a notion of (maximal) propagation velocity.  At the end of the section, I will turn to a proposal by \citet{Geroch2} and \citet{Earman} to the effect that the sense of propagation velocity given by Geroch's analysis is the salient one in connection with relativity theory.  I will then observe a close connection between the Geroch-Earman sense of superluminal velocity and Sommerfeld's wavefront velocity: indeed, as I suggested above, one might take the Geroch-Earman analysis to be a precise recovery, justification, and generalization of Sommerfeld's notion of wavefront propagation.

Geroch's analysis begins with a smooth, four dimensional manifold $M$.\footnote{We assume that the manifold is connected, paracompact, and Hausdorff.  In the next section we will return to Minkowski spacetime; here, we work in this more general setting.}  The manifold is meant to be interpreted as the spacetime manifold, though no metric need be presupposed.  One then considers ``fields'' on the manifold, understood as local sections of arbitrary fiber bundles over $M$.  That is, let $B\xrightarrow{\pi} M$ be a \emph{(smooth) fiber bundle}, which consists of a smooth surjective submersion $\pi$ between smooth manifolds $B$ and $M$ that together have the property that there exists some manifold $F$, called the typical fiber, such that for any $p\in M$, there exists an open neighborhood $U\subseteq M$ containing $p$ and a diffeomorphism $\zeta: U\times F\rightarrow \pi^{-1}[U]$ such that $\pi\circ\zeta:(q,f)\mapsto q$ for all $(q,f)\in U\times F$.  One may think of $B$, called the ``total space'', as a manifold consisting of copies of $F$ associated with each point of $M$, the ``base space''.  The map $\pi$ is called the ``projection map''; it takes a point in $B$ to the point in $M$ that lies ``beneath'' it.  The collection of all points of $B$ associated with any point $p$ of $M$, $\pi^{-1}[p]$, which forms an embedded submanifold diffeomorphic to $F$, is called the fiber at $p$.  A field, then, will be a section of this bundle, which is a smooth map $\phi:U\subseteq M\rightarrow B$, where $U$ is any smooth embedded submanifold of $M$, with the property that $\pi\circ\phi$ is the identity on $U$.

We are interested in fiber bundles whose typical fibers may be interpreted as possible physical states at a spacetime point $p$.  For instance, in electromagnetism, the case we will presently focus on, we consider a fiber bundle $B\xrightarrow{\pi} M$ whose typical fiber $F$ is the six dimensional vector space of antisymmetric rank 2 tensors $F_{ab}$ on $M$.\footnote{To see why this space is six dimensional, note that an anti-symmetric $4\times 4$ matrix has six independent elements.}  This fiber, which is naturally understood as diffeomorphic to $\mathbb{R}^6$, represents all possible values that the electromagnetic field $F_{ab}$ might take at a point $p$.  The projection map $\pi$ takes possible field values at $p$ to $p$.  A section may be interpreted as a smoothly varying assignment of field values to points of some submanifold of $M$---precisely what one would otherwise think of as an electromagnetic field.

Following Geroch, we will use the following notation, which should be understood in the general context of the abstract index notation.\footnote{Again, see the references in fn. \ref{AIN}.}  Vectors and tensors at a point in the base space $M$ (and by extension, vector and tensor fields on $M$) will be denoted using lower case Latin indices, $a,b,\ldots$.  Vectors and tensors at a point in the total space $B$, meanwhile, will be denoted using lower case Greek indices, $\alpha, \beta,\ldots$.  Finally, we will use uppercase Latin indices, $A,B,\ldots$, to indicate vectors and tensors that live in (or act on) other vector spaces.

This notation is particularly useful for treating mixed-index tensors at points $x$ of $B$: that is, tensors that may be thought of as acting on (for instance) some combination of vectors and covectors at $x$ and at $\pi(x)\in M$. To give an example of such an object, consider the pushforward along $\pi$ at $x$, $(\pi_{|x})_*:T_x B\rightarrow T_{\pi(x)} M$.  One might think of this as a tensor that takes a vector $\xi^{\alpha}$ at $x$ and returns a vector $\xi^a$ at $\pi(x)$, or, equivalently, as an object that acts on a pair $\xi^{\alpha},\eta_a$, where $\xi^{\alpha}$ is a vector at $x$ and $\eta_a$ is a covector at $\pi(x)$, to yield a real number.  In the present notation, then, this map may be written (again following Geroch) as $(\nabla\pi)^{a}{}_{\alpha}$.\footnote{The ``$\nabla$'' appearing here is not a derivative operator on either $M$ or $B$; it is used simply to invoke the fact that the pushforward map may be conceived as the differential of a smooth map between manifolds.  Note, too, that although I have defined $(\nabla\pi)^{a}{}_{\alpha}$ as the pushforward, it might equally well be thought to represent the pullback map.}  Similarly, given a section $\phi:U\rightarrow B$, the pushforward map at a point $p\in M$ may be written as $(\nabla\phi)^{\alpha}{}_a$, which should be understood as a tensor at the point $\phi(p)\in B$.  Note that the defining condition on a section guarantees that at any point $x\in\phi[U]$,
\[
(\nabla\pi)^a{}_{\alpha}(\nabla\phi)^{\alpha}{}_b = \delta^a{}_b.
\]

We will say that a vector $\xi^{\alpha}$ at a point $x\in B$ is \emph{vertical} if it is in the kernel of $(\nabla\pi)^a{}_{\alpha}$, i.e., if $(\nabla\pi)^a{}_{\alpha}\xi^{\alpha}=\mathbf{0}$.  In this case, we may think of the vector as tangent to the fiber at $\pi(x)$.  It will sometimes be convenient to indicate when a Greek index is vertical.  Again following Geroch's conventions, we will do so by adding a prime to the index---so, the vector $\xi^{\alpha'}$ at a point $x\in B$ would be vertical.  Note that for contravariant indices, we can always freely remove primes, since any vertical vector is a tangent vector, but we cannot add them, since not every tangent vector is vertical. Meanwhile, for covariant indices, we can always add primes, since any linear functional acting on all tangent vectors acts on vertical vectors, but we cannot remove them, since not every functional on vertical vectors uniquely extends to a functional on all tangent vectors.

Returning to electromagnetism, recall that the typical fiber is diffeomorphic to $\mathbb{R}^6$.  Thus the tangent space at any point of the fiber is also isomorphic to $\mathbb{R}^6$, as a vector space, and indeed, there is a canonical isomorphism between the tangent space at any point of the typical fiber and the typical fiber itself.  Composing these isomorphisms, and using the fact that the fiber at any point of the base space is diffeomorphic to the typical fiber, provides a natural sense in which any vertical vector at a point $x\in B$, i.e., any vector tangent to the fiber at $\pi(x)$, may be canonically associated with a point in $F$.  Thus a vertical vector $\delta\phi^{\alpha'}$ at a point $x\in B$ may always be thought of as an antisymmetric rank 2 tensor $\delta F_{ab}$ at $\pi(x)$.\footnote{To be clear, the $\delta$ here is part of the name of the fields $\delta\phi^{\alpha'}$ and $\delta F_{ab}$, and not a symbol of differentiation or variation.  The notation follows \citet{Geroch1}.  The idea is that the vector space structure of the fibers allows one to think of vertical vectors as differences between possible field values at a point of $M$, and the $\delta$ is meant to indicate that.}

In this language, we can now write down a general system of first-order, quasilinear differential equations on sections (i.e., fields in our general setting).  Let $k_A{}^{m}{}_{\alpha}$ and $I_A$ be smooth fields on $B$, where the $A$ index should be understood as indicating action on vectors in the ``space of equations,'' the dimension of which corresponds to the number of independent equations in the system.  Further, let $\phi:U\rightarrow B$ be a smooth section of $\pi$ (where we assume, now, that $U$ is an open subset of $M$).  Now consider the following (system of) differential equation(s):
\begin{equation}\label{Geroch-PDE}
k_A{}^{m}{}_{\alpha}(\nabla\phi)_m{}^{\alpha} + I_A=\mathbf{0}.
\end{equation}
Eq. \eqref{Geroch-PDE} should be understood to hold at each point $p\in U$, with $k$ and $I$ evaluated at $\phi(p)\in B$.  This is a first-order differential equation in $\phi$ in the sense that $(\nabla\phi)_m{}^{\alpha}$ may be understood as a generalized first-order derivative of the section.  To see this interpretation, note that for any vector $\xi^a$ at a point $p\in U$, $(\nabla\phi)^{\alpha}{}_m\xi^m=(\phi_{|p})_*(\xi^a)$ is the vector at $\phi(p)$ representing the infinitesimal direction of change along the section in ``field space'' corresponding to an infinitesimal change in the base space $M$ in the direction $\xi^a$.  And Eq. \eqref{Geroch-PDE} is quasilinear in the sense that $k_A{}^m{}_{\alpha}$ is understood to be a tensor acting on this derivative of $\phi$.  The field $I_A$, meanwhile, may be understood as the inhomogeneous, or source, term in the differential equation.

To continue developing the salient example of electromagnetism, we will now translate Eqs. \eqref{Maxwell1} into the form of Eq. \eqref{Geroch-PDE}.  In this case, the space of equations is eight dimensional, corresponding to the eight linearly independent components of Eqs. \eqref{Maxwell1};\footnote{To see this, note that on a four dimensional manifold, vectors and antisymmetric rank 3 tensors are each specified by four independent components.} this means that the capital Latin indices $A,B,\ldots$ label membership in an eight dimensional vector space.  A typical vector in this space would be of the form $\sigma^A=(s^a,s^{abc})$, where $s^{abc}=s^{[abc]}$ is totally antisymmetric; the two terms of this pair correspond to the possible coefficients of Eq. \eqref{inhomo} and Eq. \eqref{closed}, respectively.  The $I_A$ covector, meanwhile, has the form $(i_a,0)$, corresponding to the source terms in Eqs. \eqref{Maxwell1}.\footnote{For reasons that will be clear presently, one may not suppose, yet, that $i_a$ is equal to the charge-current density $J_a$ already defined.}  The most convenient way of defining $k_A{}^m{}_{\alpha}$ is in terms of its action on typical vectors and covectors in the spaces on which it acts.  So, at any point $x\in B$, and for arbitrary (co)vectors $\sigma^A=(s^a,s^{abc})$, $n_m$, and $\delta\phi^{\alpha'}=\delta F_{ab}$, at $x$, $\pi(x)$, and $x$, respectively,
\begin{equation}\label{Maxwell-G}
k_A{}^m{}_{\alpha}\sigma^An_m\delta\phi^{\alpha'}=s^b(n_mg^{ma}\delta F_{ab}) + s^{abc}(n_{[a}\delta F_{bc]}).
\end{equation}
Note that in Eq. \eqref{Maxwell-G}, we \emph{do} assume that there is a background metric $g_{ab}$ on $M$ (not necessarily the Minkowski metric), so that the base space over which the bundle of electromagnetic field values is defined is a relativistic spacetime in the ordinary sense.  This is the first time a metric is presupposed, and it is important that it appears here only to specify the system of differential equations under consideration.

To make the relationship between Eq. \eqref{Geroch-PDE} and Eqs. \eqref{Maxwell1} more transparent, note that Eq. \eqref{Maxwell-G} specifies only how $k_A{}^m{}_{\alpha}$ acts on \emph{vertical} vectors at any point of $B$, as can be seen from the $'$ on $\delta\phi^{\alpha'}$.  The field $k_A{}^m{}_{\alpha}$, meanwhile, was defined as an object that acts on all vectors at a point of $B$.  In fact, though, it is only the ``vertical part'' of $k_A{}^m{}_{\alpha}$ (in the last index) that involves the derivatives of the fields.  Thus there is some freedom in how we write $k_A{}^m{}_{\alpha}$; this freedom is off-set by a corresponding freedom in $I_A$, required to leave Eq. \eqref{Geroch-PDE} invariant.\footnote{See \citet[pp. 8--9]{Geroch1} for more details.  The freedom amounts to a choice of ``linear connection'' on the bundle of field values.}  In general, however, we may always choose $k_A{}^m{}_{\alpha}$ in such a way that the derivatives of the sections are taken relative to the covariant derivative operator $\nabla$ compatible with the metric $g_{ab}$.  When we do so, the interaction vector $I_A$ becomes $I_A=(J_a,0)$, where $J_a$ is the standard charge-current density on $M$.  With these choices, we may write $k_A{}^m{}_{\alpha}=(\delta^y{}_ag^{mx},\epsilon_{abcp}\epsilon^{pmxy})$, where each term acts on $\nabla_mF_{xy}$, understood as a field on spacetime.

Up to this point in the section, we have merely developed a general formalism for treating partial differential equations and then translated Maxwell's equations into that formalism.  Now comes the pay-off.  In the following definitions and propositions, we will define the notion of propagation velocity we have been working towards.  Once we have given the general statement of the necessary propositions, we will return to electromagnetism once again.

Suppose we are given a differential equation in the form of Eq. \eqref{Geroch-PDE}.  A \emph{hyperbolization} is a smooth field $h^A{}_{\beta'}$ on $B$ such that (1) the field $h^A{}_{\beta'}k_A{}^m{}_{\alpha'}$ is symmetric in $\alpha'$, $\beta'$ and (2) at each point $x\in B$, there exists a covector $n_m$ at $\pi(x)$ such that $h^A{}_{\beta'}k_A{}^m{}_{\alpha'}n_m$ is positive definite, i.e., is such that for all non-zero vertical vectors $\xi^{\alpha'}$ at $x$, $n_mh^A{}_{\beta'}k_A{}^m{}_{\alpha'}\xi^{\alpha'}\xi^{\beta'}>0$.  We call such a field a hyperbolization because the differential equation
\begin{equation}\label{Geroch-PDE-hyper}
h^A{}_{\beta'}\left(k_A{}^{m}{}_{\alpha}(\nabla\phi)_m{}^{\alpha} + j_A\right)=\mathbf{0}
\end{equation}
is symmetric hyperbolic,\footnote{This means, roughly, that it is a differential equation admitting wave-like solutions.  For more precise characterizations, see, for instance, \citet[\S 7.3]{Evans} or \citet[Ch. 2]{Lax}.} and any solution to Eq. \eqref{Geroch-PDE} is also a solution to Eq. \eqref{Geroch-PDE-hyper}.\footnote{Of course, ultimately we care about the converse, i.e., when solutions of Eq. \eqref{Geroch-PDE-hyper} are solution to Eq. \eqref{Geroch-PDE}.  But addressing this issue would require a discussion of constraints, which is not necessary for the arguments of the present paper.  See \citet[\S 4]{Geroch1} for a general discussion.}

Now suppose one has a differential equation in the form of Eq. \eqref{Geroch-PDE}, and suppose it admits a hyperbolization $h^A{}_{\beta'}$.  Then at each point $x\in B$, let $s_x$ be the collection of covectors $n_a$ at $\pi(x)$ such that $n_mh^A{}_{\beta'}k_A{}^m{}_{\alpha'}$ is positive definite.  The set $s_x$ will in general be an open convex cone of covectors at $\pi(x)$.  Now let $C_x$ be the collection of vectors $\xi^a$ at $\pi(x)$ with the property that $\xi^an_a\geq0$ for every covector $n_a\in s_x$.  We call $C_x$, which is a closed convex cone of vectors at $\pi(x)$, the \emph{causal cone} at $\pi(x)$ (for field value $x$); vectors in $C_x$ will be called causal vectors at $x$.  It is crucial to emphasize that this causal cone is defined \emph{without reference to a spacetime metric}.  Of course, Lorentz-signature metrics are also associated with causal cones; in what follows, we will distinguish causal cones, which are associated with a system of differential equations, from ``metric lightcones'', which are the (causal) cones associated with a background spacetime metric.  In general, there should be no expectation that these will coincide.  Similarly, we will attempt to clearly distinguish causal vectors in the sense of elements of $C_x$ for some differential equation from causal vectors in the standard sense of timelike or null vectors relative to a metric.

In general, the causal cone associated with a system of equations has the following interpretation: it is the collection of ``signal propagation directions,'' or perhaps better, ``signal propagation 4-velocities'' for the field at $\pi(x)$.  To justify this interpretation, consider the following definitions and proposition.  Let $S$ be a three dimensional embedded submanifold of $M$, and suppose there are fields (i.e., local sections) $\psi:S\rightarrow B$ on $S$.  Then we will say that $(S,\psi)$ is \emph{initial data} for our system of differential equations \eqref{Geroch-PDE} if at every point $p$ of $S$, the normals $n_a$ to $S$ at $p$ are elements of $s_{\psi(p)}$.  The idea is that initial data is a specification of field values on a collection of points that are not ``causally related'' according to the standard given by $C_x$.  A \emph{solution} $(U,\phi)$ to the differential equation for initial data $(S,\psi)$ is a neighborhood $U\subseteq M$ containing $S$ and fields $\phi:U\rightarrow B$ on $U$ such that $\phi$ satisfies the system of equations and $\phi_{|S}=\psi$.  Given a solution $(U,\phi)$, we will say a smooth curve $\gamma:I\rightarrow M$ is \emph{causal relative to $(U,\phi)$} if its tangent vector at each point of its image, $\vec{\gamma}_{|\gamma(s)}$, is an element of $C_{\phi(\gamma(s))}$ whenever $\gamma(s)\in U$.  A point $p\in U$ is an \emph{endpoint} of a smooth curve $\gamma:I\rightarrow M$ that is causal relative to some solution $(U,\phi)$ if $p$ is such that, for any open set $O$ containing $p$, there is a parameter $s_0\in I$ such that either for all $s\geq s_0$ or all $s\leq s_0$, $\gamma(s)\in O$.  Finally, given initial data $(S,\psi)$, the \emph{domain of dependence} of $S$, $D(S)$, is the collection of all points $p\in M$ such that (1) there exists a solution $(U,\phi)$ for initial data $(S,\psi)$ where $p\in U$, and (2) given any smooth curve $\gamma:I\rightarrow M$ whose image contains $p$, if $\gamma$ is causal relative to a solution $(U,\phi)$ such that $p\in U$, and $\gamma$ is without endpoint in $U$, then $\gamma[I]$ intersects $S$.

Given these definitions, the following is a consequence of the basic uniqueness results for symmetric hyperbolic systems.\footnote{See \citet[Appendix B]{Geroch1}.  Of course, there are general existence results, too, though these require a treatment of constraints.  For our purposes, all that matters is uniqueness.}
\begin{prop}Let $(S,\psi)$ and $(S,\psi{}')$ be initial data for a differential equation of the form of Eq. \eqref{Geroch-PDE} with fixed hyperbolization.  Suppose that there is some open (in the submanifold topology on $S$) subset $A\subset S$ on which $\psi$ and $\psi{}'$ agree.  Then any solutions $(U,\phi)$ for initial data $(S,\psi)$ and $(U',\phi')$ for initial data $(S,\psi{}')$ must agree on the domain of dependence $D(A)$, i.e., $\phi_{|p}=\phi'_{|p}$ for all points $p\in U\cap U'\cap D(A)$.\end{prop}
In other words, solutions on $D(A)$ are entirely fixed by the initial data on $A$.  This means that any initial data off of $A$ cannot contribute to the solution on $D(A)$, and thus, one could not perturb the initial data off of $A$ in such a way as to send a signal (say) or otherwise affect field values within $D(A)$.  It is in this sense that the causal cones at a point determine the possible signal propagation directions.

We can now return to electromagnetism, to consider the causal cones associated with Maxwell's equations.  In that case, the general hyperbolization $h^A{}_{\beta'}$ at a point $x\in B$ may be defined in terms of its action on an arbitrary vertical vector $\delta\phi^{\alpha'}=\delta F_{ab}$, as
\begin{equation}
h^A{}_{\alpha'}\delta\phi^{\alpha'}=(\delta F^a{}_{m} \zeta^m, \frac{3}{2}\zeta^{[a}\delta F^{bc]}),
\end{equation}
where all indices are raised with $g_{ab}$, and where $\zeta^a$ is some timelike vector at $\pi(x)$.

Contracting the $A$ index on $h^A{}_{\alpha'}$ with $k_A{}^m{}_{\alpha'}$ as defined in Eq. \eqref{Maxwell-G} yields, for arbitrary vertical vectors $\delta\phi^{\alpha'}=\delta F_{ab}$ and $\delta\hat{\phi}^{\alpha'}=\delta\hat{F}_{ab}$ and arbitrary covector $n_m$,
\begin{align}
h^A{}_{\alpha'}k_A{}^m{}_{\beta'}n_m\delta\phi^{\alpha'}\delta\hat{\phi}{}^{\beta'}&=\delta F^b{}_{m}\delta \hat{F}_{ab} \zeta^m n^a +\frac{3}{2}\zeta^{[a}\delta F^{bc]}(n_{[a}\delta \hat{F}_{bc]})\\
&=2\zeta^a n^b\left(\delta F^m{}_{(a}\delta \hat{F}_{b)m}+\frac{1}{4}g_{ab}\delta F_{mn}\delta \hat{F}^{mn}\right).\label{EM-hyper}
\end{align}
This last expression allows us to confirm that $h^A{}_{\alpha'}$ as just defined is a hyperbolization, since (1) Eq. \eqref{EM-hyper} is manifestly invariant under exchange of $\delta F_{ab}$ and $\delta\hat{F}_{ab}$ (which means that $h^A{}_{\alpha'}k_A{}^m{}_{\beta'}$ is symmetric in $\alpha'$ and $\beta'$) and (2) $h^A{}_{\alpha'}k_A{}^m{}_{\beta'}n_m$ is positive definite on vertical vectors if (and only if) $n_a$ is timelike and co-oriented with $\zeta^a$ (relative to the background spacetime metric, $g_{ab}$), since in this case Eq. \eqref{EM-hyper} is positive whenever $\delta F_{ab}=\delta\hat{F}_{ab}$.\footnote{To see that (2) holds, note that $\zeta^a\left(\delta F^m{}_{(a}\delta F_{b)m}+\frac{1}{4}g_{ab}\delta F_{mn}\delta F^{mn}\right)=T_{a}{}^b\zeta^a$, where $T^{ab}$ is the energy momentum tensor associated with the electromagnetic field $\delta F_{ab}$. Since the energy-momentum tensor associated with \emph{any} electromagnetic field satisfies the Dominant Energy Condition, $T_{a}{}^b\zeta^a$ is causal and co-oriented with $\zeta^a$ \citep[see ][\S 2.6]{MalamentGR}.  Thus $T_{a}{}^{b}\zeta^a n_b >0$ for any timelike $n_b$ co-oriented with $\zeta^a$.  This establishes the ``if'' clause.  For the converse, note that by varying $\delta F_{ab}$ (for fixed $\zeta^a$), one can force $T_{a}{}^b\zeta^a$ to be \emph{any} causal vector co-oriented with $\zeta^a$.  Thus, for any null or spacelike $n^a$, one can always choose $\delta F_{ab}$ so that $\zeta^an_bT_{a}{}^{b}=\zeta^an^b\left(\delta F^m{}_{(a}\delta F_{b)m}+\frac{1}{4}g_{ab}\delta F_{mn}\delta F^{mn}\right)\leq 0$.  This establishes the ``only if'' direction.}   This last observation (or rather, its converse) implies that for any point $x\in B$, $s_x$ consists in precisely the timelike covectors at $\pi(x)$.  Thus, for electromagnetism, the causal cone at any point $x$ is precisely the collection of causal vectors (in the ordinary sense of timelike or null vectors, relative to $g_{ab}$) co-oriented with $\zeta^a$, the timelike vector determining the hyperbolization.  It follows that for electromagnetism, the possible signal propagations are precisely the causal vectors (relative to a metric $g_{ab}$).

The discussion thus far has concerned the \emph{possible} signal propagation velocities associated with a field.  But in fact, one can say a bit more.  Again with a hyperbolization fixed for some differential equation in the form of Eq. \eqref{Geroch-PDE}, call the (non-zero) boundary of $s_x$, i.e., the collection of non-zero covectors at $\pi(x)$ that lie in the closure of $s_x$ but not the interior of $s_x$, the \emph{characteristic covectors} at $\pi(x)$ (for field value $x$).  Suppose that one has a solution $(U,\phi)$ of such a differential equation (for some initial data or other).  A \emph{characteristic surface} for that solution is a three dimensional embedded submanifold $\Sigma\subset U$ that has a characteristic covector as a normal at every point $p\in\Sigma$.

Characteristic surfaces are of interest because of the following classical result of the theory of hyperbolic partial differential equations:\footnote{For more on the relationship between characteristics and causal cones, in terms that are quite close to those developed by \citet{Geroch1}, see \citet{Khavkine}.}
\begin{prop}[\citet{Courant+Lax}] Suppose $(S,\psi)$ is initial data for a differential equation of the form of Eq. \eqref{Geroch-PDE} with fixed hyperbolization, and suppose that $\psi$ is smooth everywhere except across a two dimensional embedded submanifold $\Gamma\subseteq S$, where $\psi$ or one of its derivatives is assumed to exhibit a jump discontinuity.  Then any solution $(U,\phi)$ for initial data $(S,\psi)$ is smooth everywhere except across the characteristic surfaces $\Sigma\subseteq U$ containing $\Gamma$; across these surfaces, $\phi$ or one of its derivatives exhibits a jump discontinuity.\end{prop}
To interpret this proposition, note that the tangents to characteristic surfaces will be causal in the sense that, at each point of these surfaces, one can always find a basis of three vectors, all of which are elements of $C_x$.  Moreover, at each point, these causal tangent vectors will be boundary points of $C_x$.  This means that, given our interpretation of causal cones, the tangents to characteristic surfaces are, in a precise sense, the \emph{maximal} 4-velocities associated with the propagation of a field governed by the given equation.  Thus the moral of the proposition is that \emph{discontinuities} in initial data (or its derivatives) propagate along characteristic surfaces, i.e., with ``maximal'' 4-velocities.

On the basis of the considerations offered above, \citet{Geroch2} and \citet{Earman} have argued that the salient sense of (maximal) propagation velocity for a field is the one given by the causal cones associated with that field, in the sense just described.  One might condense their discussion into a criterion of superluminal propagation for fields governed by some differential equation as follows.
\begin{con}[Geroch-Earman] Suppose one is given a system of differential equations of the form of Eq. \eqref{Geroch-PDE} on sections of a bundle $B\xrightarrow{\pi} M$ of possible field values over a spacetime $(M,g_{ab})$, and suppose one has a fixed hyperbolization of the system.  Then solutions to this system of equations may propagate superluminally if and only if there exists an open set $O\subseteq B$ such that for all $x\in O$, the causal cone $C_x$ contains as a proper subset one lobe of the metric lightcone associated with $g_{ab}$ at $\pi(x)$.\end{con}
The discussion above shows that when this condition is satisfied, for at least some initial data, discontinuities in that data will propagate outside the metric lightcone. This is the sense in which such field theories exhibit superluminal propagation.  Of course, one may also readily cast this condition as a sufficient criterion for \emph{no} superluminal propagation, by restricting attention to systems of equations that do not meet the stated condition.\footnote{One might also take this to be a necessary condition for no superluminal propagation.  But that would require further argument.  The condition states that a field \emph{may} propagate superluminally if the causal cone $C_x$ contains a lobe of the metric lightcone, not that fields \emph{do} so propagate.  Just consider: the causal cones associated with a fluid correspond to the ``sound cone''.  But it hardly follows that wind always travels at the speed of sound!  In other words, not all salient features of a field correspond to discontinuities in a field or its derivatives, and thus they need not follow the characteristics.}

It is worth noting what Earman and Geroch take the significance of this criterion to be.  Earman argues that with this criterion of superluminal (non-)propagation in mind, other standard criteria---including the so-called dominant energy condition---are neither necessary nor sufficient for no superluminal propagation.  He also argues that Geroch's approach to characterizing superluminal propagation provides insight into how to incorporate requirements of ``relativistic causality'' into quantum field theory.  Perhaps more importantly for the present discussion, Geroch, argues (and Earman appears to accept) that on this understanding of propagation velocity, \emph{superluminal propagation presents no contradiction with relativity}.  The idea is that relativity, be it special or general, is merely one system of differential equations on a manifold.  Solutions to these equations have causal cones corresponding to the metric lightcones, but this does not constrain the causal cones of other fields.  As long as a field has a well-defined initial value formulation, Geroch argues, it is perfectly consistent with relativity.\footnote{\label{worries} Geroch also argues that no other inconsistencies---such as the ``grandfather paradox''---arise with such theories.  We will not rehearse those arguments here; the short version is that a well-posed initial value formulation guarantees self-consistency of solutions, so that a field will never evolve so as to change its initial data.  A further worry about Geroch's claim that superluminal propagation is consistent with relativity is that Einstein's equation plays no role in the Geroch-Earman condition.  I will briefly return to this point at the end of section \ref{woah}.}

I am now in a position to make the promised connection between the Geroch-Earman criterion for superluminal propagation and Sommerfeld's criterion.  Recall that Sommerfeld identified the propagation of a \emph{wavefront}, which he defined as a jump discontinuity in a solution to Maxwell's equations, as the salient sense of field propagation in relativity theory.  Given the interpretation of characteristic surfaces as the surfaces across which a solution (or the derivatives of a solution) to some hyperbolic system of differential equations may have a jump discontinuity, and the relationship between causal cones and characteristic surfaces, we can now see that Sommerfeld's wavefront velocity corresponds precisely to (a special case of) the propagation velocity given by Geroch and Earman.  And indeed, as we have seen here, in the case of electromagnetic fields in a vacuum, i.e., solutions of Maxwell's equations, the (causal) tangents to the characteristic surfaces are precisely the null vectors, just as Sommerfeld argued.  In this sense, the discussion above amounts to yet another argument that the relevant propagation velocity, by which we now mean the (suitably generalized) wavefront velocity, for electromagnetic fields in a vacuum is $c$.  Of course, Sommerfeld also argued that the wavefront velocity would be $c$ in \emph{any} medium.  It is to this claim that we turn in the next two sections.

\section{Dielectrics Revisited}\label{Gordon}

With the Geroch-Earman criterion for superluminal propagation in hand, we can now return to the question with which we began, concerning whether under some circumstances, the propagation of an electromagnetic field in a dielectric medium is properly conceived as superluminal.  To address the question in the present context, we require a system of differential equations governing the propagation of electromagnetic fields in a medium.  There is a standard choice here, known as the ``macroscopic'' Maxwell equations.\footnote{See, for instance, \citet[Ch. 9]{Landau+Lifshitz} or \citet[Ch. 6]{Jackson}.}  These are most familiarly written relative to a choice of constant timelike vector field $\xi^a$, analogously to Eqs. \ref{Maxwell2}.  Once again limiting attention to Minkowski spacetime $(M,\eta_{ab})$, the equations are:\footnote{Of course, one could take yet another step back from a covariant four dimensional presentation of the equations, along the lines of Eqs. \ref{Maxwell3}.  Indeed, it is in this form that the equations are presented in classical references, such as those cited in the previous footnote.}
\begin{subequations}\label{MacroMaxwell2}
\begin{align}
\partial_b B^b &= 0 \\
 \epsilon^{abc} \partial_b E_c &= -\xi^n\nabla_b B^a  \\
\partial_b D^b &= \sigma^{\text{ext}} \\
\epsilon^{abc} \partial_b H_c &= \xi^n\nabla_n D^a + (J^{\text{ext}})^a.
\end{align}
\end{subequations}
Here we have introduced four new fields: $D^a$, the electric displacement field, $H^a$, the magnetic field,\footnote{To avoid notational conflicts with the literature, I am following standard practice and calling $H^a$ the magnetic field; $B^a$, in this context, is then called the ``magnetic induction''.  The usage is confusing, however, since $B^a$ continues to represent the ``averaged'' or ``macroscopic'' incident magnetic field (just as $E^a$ represents the ``macroscopic'' electric field), and $H^a$ characterizes the response of the medium.  See, for instance, \citet[pp. 106-7]{Landau+Lifshitz} or \citet[pp. 13-4]{Jackson}.} and $\sigma^{\text{ext}}$ and $J^{\text{ext}}$, which are the ``external'' or ``free'' charge and 3-current densities, i.e., the charge and 3-current densities not associated with the medium.  The $D^a$ and $H^a$ fields characterize the electromagnetic response of the medium; we will presently make a (fairly standard) assumption regarding their relationship to the incident fields $E^a$ and $B^a$.  It is worth emphasizing that, as with $E^a$ and $B^a$, $D^a$ and $H^a$ are defined relative to the constant timelike vector field $\xi^a$ (recall section \ref{prelim}).  There is a sense, however, in which there is now a privileged choice of observer field, since the medium has an associated 4-velocity, and so we assume that $\xi^a$ is the 4-velocity of the medium.\footnote{\label{momentum}  To be clear about the role that $\xi^a$ plays here: at any point, given any timelike (relative to the spacetime metric) vector $\xi^a$, one can define $E^a$, $B^a$, $D^a$, and $H^a$ relative to $\xi^a$.  In order to write Eqs. \eqref{MacroMaxwell2}, however, one requires $\xi^a$ to be constant on an open neighborhood of $p$.  Meanwhile, in order to understand the relationship between $E^a$ and $D^a$, and $B^a$ and $H^a$, in the standard way, as in Eqs. \eqref{const1} and \eqref{const2} below, it is necessary to take the $4-$velocity of the medium to define a privileged observer field, since $D^a$ and $H^a$ represent the response on the medium, in its own frame, to the incident electromagnetic fields.  But in order to associate the field $\xi^a$ relative to which Eqs. \eqref{MacroMaxwell2} are written with the 4-velocity of a medium we must assume that the $4-$velocity of the medium is constant, which of course is a very strong assumption.  Indeed, one might worry that the interaction of the electromagnetic field and the medium would itself produce acceleration in the medium.  But in order to treat such acceleration, one would require a detailed theory of the medium, with additional differential equations governing the vector field $\xi^a$, at least.  This may be a reason to reject the macroscopic Maxwell equations altogether, at least for foundational purposes.  Conversely, insofar as we take the macroscopic Maxwell equations, as expressed by Eqs. \eqref{MacroMaxwell2}, to be the correct system of equations for an electromagnetic field in a medium, we are apparently forced to this assumption, and indeed, we will adopt it here.  All that said, one can relax the assumption once one moves to Eqs. \eqref{MacroMaxwell3}, below, which are at least well-defined even in the case of curved spacetime, where constant vector fields in general do not exist.  Even in that context, however, the macroscopic Maxwell equations provide no insight into the acceleration of the medium in response to the incident fields.}

For present purposes, it will be convenient to re-write Eqs. \eqref{MacroMaxwell2} in a form analogous to Eqs. \eqref{Maxwell1}.  To do so, we first define two antisymmetric tensors:
\begin{subequations}\label{MacroFP}
\begin{align}
F_{ab} &= 2E_{[a}\xi_{b]} + \epsilon_{ab}{}^{cd}\xi_c B_d \label{Fdef}\\
P_{ab} &= 2D_{[a}\xi_{b]} + \epsilon_{ab}{}^{cd}\xi_c H_d \label{Pdef}.
\end{align}
\end{subequations}
These fields satisfy the following differential equations.
\begin{subequations}\label{MacroMaxwell3}
\begin{align}
\nabla_a P^{ab}&=(J^{\text{ext}})^b\label{Max1}\\
\nabla_{[a}F_{bc]}&=\mathbf{0}\label{Max2}
\end{align}
\end{subequations}
where $(J^{\text{ext}})^b$ is the ``external'' charge-current density.  In what follows, we will assume that $(J^{\text{ext}})^b=\mathbf{0}$.

We assume the medium is linear, which means that the electromagnetic properties of the medium are characterized by the following constitutive relations:
\begin{align}
D_a &= \varepsilon_a{}^b E_b \label{const1}\\
B_a &= \mu_a{}^b H_b \label{const2}
\end{align}
where $\varepsilon_a{}^b$ is the \emph{electric permittivity} and $\mu_a{}^b$ is the \emph{magnetic permeability}.  In principle, these may be arbitrary tensors (for anisotropic media) depending on a number of parameters, including location in spacetime (for inhomogeneous media) and frequency (for dispersive media).\footnote{There is a puzzle concerning how to think about ``dispersive media'' in curved spacetime, where Fourier transforms are not generally well-defined.  But we set this issue aside, since we are working in Minkowski spacetime.}  One may also consider non-linear media, where $\varepsilon$ and $\mu$ may depend on the field strengths and where there may be other terms in Eqs. \eqref{const1} and \eqref{const2}.  In what follows, however, we will consider just the very simplest case, where $\varepsilon_a{}^b = \varepsilon\delta_a{}^b$ and $\mu_a{}^b=\mu\delta_a{}^b$, for constant scalar fields $\varepsilon$ and $\mu$.  This corresponds to the case where the medium's response to an incident beam is homogeneous, isotropic, and non-dispersive.\footnote{Note that the assumption that the medium is non-dispersive means that the phase velocity and group velocity coincide.  (Recall the discussion surrounding Eq. \eqref{groupv2}.)  Note, too, that I mean to assume only that the medium's response is homogeneous and isotropic in the sense of the stated condition on the tensors $\varepsilon_a{}^{b}$ and $\mu_a{}^b$.  I do not mean to assume that the medium is homogeneous or isotropic in other senses, and in particular, I do not mean to make any assumptions about the energy-momentum tensor associated with the medium.}

In this regime, we can rewrite Eqs. \eqref{MacroFP} in terms of just the electric and magnetic fields, $E^a$ and $H^a$:
\begin{subequations}
\begin{align}
F_{ab} &= 2E_{[a}\xi_{b]} + \mu \epsilon_{ab}{}^{cd}\xi_c H_d \\
P_{ab} &= 2\varepsilon E_{[a}\xi_{b]} + \epsilon_{ab}{}^{cd}\xi_c H_d \label{Prewrite1}.
\end{align}
\end{subequations}
Moreover, we can write the electric and magnetic fields relative to $\xi^a$ in terms of $\xi^a$ and $F_{ab}$ as:
\begin{align}
E^a&=F^a{}_{b}\xi^b\\
H^a&=\frac{1}{2\mu} \epsilon^{abcd}\xi_b F_{cd}.
\end{align}
Plugging these into Eq. \eqref{Prewrite1} allows us to write $P_{ab}$ in terms of $F_{ab}$, as:
\begin{equation}
P_{ab}=\frac{1}{\mu}\left(2(1-n^2)\xi_{[a}F_{b]n}\xi^n + F_{ab}\right)\label{Prewrite2}.
\end{equation}
Here we have defined the index of refraction, $n$, as $n=\sqrt{\varepsilon\mu}$.\footnote{Why is this the index of refraction?  In short, because if we rewrite Eqs. \eqref{MacroMaxwell2} using Eqs. \eqref{const1} and \eqref{const2} under the present assumptions, and then consider wavelike solutions, $\sqrt{\varepsilon\mu}$ plays precisely the role one would expect of $n$.}   Since we have assumed that $\varepsilon$ and $\mu$ are constant scalar fields, $n$ is also a constant scalar field.  In particular, $n$ is not a function of frequency.

Now, appealing to the previous assumption that $\xi^a$ is constant (and thus, $\nabla_a\xi^b=\mathbf{0}$), we can write the divergence of $P^{ab}$ in terms of $F_{ab}$ as:
\begin{equation}
0=\nabla_aP^{ab}=\frac{1}{\mu}\left(2(1-n^2)\xi^n\xi^{[a}\nabla_aF^{b]}{}_n + \nabla_aF^{ab}\right)
\end{equation}
Thus, at least under the present assumptions (including $(J^{\text{ext}})^b=\mathbf{0}$), we can think of the macroscopic Maxwell equations (Eqs. \eqref{Max1} and \eqref{Max2}) as:
\begin{subequations}
\begin{align}
2(1-n^2)\xi^n\xi^{[a}\nabla_aF^{b]}{}_n + \nabla_aF^{ab}&=\mathbf{0}\label{Max1new}\\
\nabla_{[a}F_{bc]}&=\mathbf{0}\label{Max2new}.
\end{align}
\end{subequations}

Eqs. \eqref{Max1new} and \eqref{Max2new} constitute a system of quasi-linear, first-order partial differential equations, which means they are amenable to the Geroch analysis.  This analysis is simplified by the observation that these equations may be rewritten in terms of an effective metric $\tilde{\eta}_{ab}$ (known as ``Gordon's metric'') as:\footnote{For more on Gordon's metric, see \citet{Gordon} or \citet[\S E.4]{Hehl+Obukhov}.   See also \citet{Novello+Bittencourt} for a discussion of the relationship between Gordon's metric and the propagation of discontinuities.  Note that to recover Eq. \eqref{Max1eff}, we need to raise $b$ using $\tilde{\eta}^{ab}$---or equivalently, observe that Eq. \eqref{Max1eff} simplifies if we multiply both sides by $\tilde{\eta}_{bc}$.  Of course, this simplification only works in the source-free case.  One might also wonder why no change is required for Eq. \eqref{Max2eff} to reflect the fact that we are working with an ``effective metric''. Another way of writing this equation would be as $\tilde{\epsilon}^{abcd}\nabla_bF_{cd}=\mathbf{0}$, where $\tilde{\epsilon}_{abcd}$ is the volume element associated with $\tilde{\eta}_{ab}$.  Note, though, that $\tilde{\epsilon}_{abcd}=\frac{1}{n}\epsilon_{abcd}$, and so $\tilde{\epsilon}^{abcd}\nabla_bF_{cd}=\mathbf{0} \Leftrightarrow \epsilon^{abcd}\nabla_bF_{cd}=\mathbf{0}$.}
\begin{subequations}
\begin{align}
\tilde{\eta}^{an}\nabla_aF_{nb}&=\mathbf{0}\label{Max1eff}\\
\nabla_{[a}F_{bc]}&=\mathbf{0}\label{Max2eff}.
\end{align}
\end{subequations}
where
\begin{equation}
\tilde{\eta}_{ab} = \eta_{ab} + \frac{1-n^2}{n^2}\xi_a \xi_b
\end{equation}
with inverse metric
\begin{equation}
\tilde{\eta}^{ab} = \eta^{ab} - (1-n^2)\xi^a\xi^b.
\end{equation}
Note that $\tilde{\eta}_{ab}$ is a flat, geodesically complete Lorentzian metric on the manifold $M$.  (It is flat because it is constant with respect to $\nabla$, the Levi-Civita derivative operator associated with $g_{ab}$.)  Indeed, when $n=1$, we have $\tilde{\eta}_{ab}=\eta_{ab}$.  Meanwhile, when $n>1$, $\tilde{\eta}_{ab}$ can be thought of as implementing an effective \emph{narrowing} of the lightcones relative to $\eta_{ab}$.  In other words given any vector $\zeta^a$ at a point $p$, if $\zeta^a\zeta^b\tilde{\eta}_{ab}\geq 0$, then $\zeta^a\zeta^b\eta_{ab} = \zeta^a\zeta^b\tilde{\eta}_{ab} - \frac{1-n^2}{n^2}(\xi^a\zeta_a)^2\geq 0$, so the causal vectors relative to $\tilde{\eta}_{ab}$ at any point are a subset of the causal vectors relative to $\eta_{ab}$ at that point.  Conversely, if $n<1$, $\tilde{\eta}_{ab}$ implements an effective \emph{widening} of the lightcones.  Eqs. \eqref{Max1eff} and \eqref{Max2eff}, then, are none other than Maxwell's equations in Minkowski spacetime---but with Minkowski metric $\tilde{\eta}_{ab}$, rather than $\eta_{ab}$.

Given this expression of the system of equations, the analysis of vacuum electromagnetism in the previous section carries over intact---with $\tilde{\eta}_{ab}$ taking the place of $\eta_{ab}$.  Once again, the fibers of the bundle $B\xrightarrow{\pi} M$ are precisely as in standard electromagnetism: they consist of the six dimensional vector space of anti-symmetric rank 2 tensors at a point $p\in M$.  A typical vertical vector is of the form $\delta \phi^{\alpha'}=\delta F_{ab}=\delta F_{[ab]}$.  A typical vector in the space of equations is $\sigma^A=(s^a,s^{abc})$, where $s^a$ is a vector and $s^{abc}=s^{[abc]}$ is an antisymmetric third rank tensor.  We can define the field $k_A{}^m{}_{\alpha'}$ by defining its action on typical vectors $\delta\hat{\phi}^{\alpha'}$ and $n_m$ in the relevant spaces:
\begin{equation}
k_A{}^m{}_{\alpha'}\sigma^A n_m \delta\hat{\phi}^{\alpha'} = s^a(-\tilde{\eta}^{bm}n_m\delta\hat{F}_{ab})+s^{abc}(n_{[a}\delta\hat{F}_{bc]}).
\end{equation}
The general hyperbolization $h^A{}_{\beta'}$ at a point $x\in B$ may again be defined by its contraction with an arbitrary vertical vector $\delta\phi^{\alpha'}=\delta F_{ab}$, as:
\begin{equation}
h^A{}_{\alpha'}\delta\phi^{\alpha'}=(\delta F^a{}_{m} \zeta^m, \frac{3}{2}\zeta^{[a}\delta F^{bc]}),
\end{equation}
where now all indices are raised with $\tilde{\eta}_{ab}$.  Once again, $\zeta^a$ is an arbitrary timelike vector (relative to $\tilde{\eta}_{ab}$) at $\pi(x)$.

We are now in a position to identify the causal cones associated with the macroscopic Maxwell equations.  Contracting the $A$ index on $h^A{}_{\alpha'}$ with $k_A{}^m{}_{\alpha'}$ as just defined yields, for arbitrary vertical vectors $\delta\phi^{\alpha'}=\delta F_{ab}$ and $\delta\hat{\phi}^{\alpha'}=\delta\hat{F}_{ab}$ and arbitrary covector $n_m$,
\begin{equation}
h^A{}_{\alpha'}k_A{}^m{}_{\beta'}n_m\delta\phi^{\alpha'}\delta\hat{\phi}^{\beta'}=2\xi^a n^b\left(\delta F^m{}_{(a}\delta \hat{F}_{b)m}+\frac{1}{4}\tilde{\eta}_{ab}\delta F_{mn}\delta \hat{F}^{mn}\right).\label{MEM-hyper}
\end{equation}
Eq. \eqref{MEM-hyper} allows us to determine the causal cones associated with Eqs. \eqref{MacroFP}.  Analogously to the vacuum case, Eq. \eqref{MEM-hyper} is positive definite whenever $n_m$ is timelike \emph{relative to $\tilde{\eta}_{ab}$}. This means that for any point $x\in B$, $s_x$ consists in precisely the $\tilde{\eta}$-timelike covectors at $\pi(x)$ co-oriented with $\zeta^a$, and thus $C_x$ consists of the $\tilde{\eta}$-causal vectors. It follows that when $n>1$, the causal cones associated with the macroscopic Maxwell equations are narrower than the spacetime metric lightcones.  When $n<1$, however, the causal cones become \emph{wider} than the metric lightcones.  In other words, under the assumptions of the present section, and for $n<1$, solutions to the macroscopic Maxwell equations for an electromagnetic field in a moving dielectric propagate superluminally in the Geroch-Earman sense---and thus, given the discussion at the end of the previous section, in the Sommerfeld-Brillouin sense as well.  In other words, it would seem that at least for some parameter values, electromagnetic field \emph{do} propagate superluminally in a dielectric medium.

\section{Sommerfeld Revisited}\label{Sommerfeld}

The result of the previous section---that for some parameter values, electromagnetic fields in a dielectric may propagate superluminally in the Geroch-Earman(-Brillouin-Sommerfeld) sense---may be surprising.  After all, as I noted at the end of section \ref{Milonni}, there is an oft-cited and widely accepted argument, due to \citet{Sommerfeld}, that wavefront velocities are \emph{always} $c$ in a medium, and thus that superluminal propagation of electromagnetic fields in a dielectric medium is impossible.  Since the velocity in question in the Geroch-Earman analysis is precisely the wavefront velocity, there appears to be a contradiction between Sommerfeld's argument and the results we have just seen.

To resolve the tension, one needs to look at the details of Sommerfeld's argument, which thus far I have simply cited without elaboration.  It may be presented as follows.\footnote{The version of the argument I present here follows the presentation of \citet[Ch. 7]{Oughstun+Sherman}. \citet[\S 7.11]{Jackson} offers a similar, albeit more impressionistic, version that suppresses some technical details.  (In this instance, \citet{Milonni} is misleading; the argument he presents is a variation on Jackson that, so far as I have been able to tell, cannot work.)  Sommerfeld's version of the argument amounts to a special case of Oughstun and Sherman's version; since Sommerfeld works with a specific model of the interaction between matter and the electromagnetic field, his version makes it more difficult to isolate the essential assumptions.  Nonetheless, I will continue to refer to what I present here as ``Sommerfeld's argument,'' since it originates with him both in spirit and in much substance.}  First, as in the previous section, we suppose that we have some dielectric medium filling all of Minkowski spacetime, and we have an electromagnetic field within the medium.  For simplicity, suppose we are working in two spacetime dimensions, with fixed (standard) coordinates $x$ and $t$.  Suppose, too, that the medium has a constant 2-velocity and that the coordinates have been chosen so that $\xi^a=\left(\frac{\partial}{\partial t}\right)^a$, where $\xi^a$ is the 2-velocity of the medium.  Then we can write Eq. \eqref{genSol}, the general solution to Maxwell's equations, as:
\begin{equation}\label{genSol2}
E(x,t) = \frac{1}{\sqrt{2\pi}}\int_{\mathbb{R}} d k C(k) e^{i(kx-\omega(k) t)}.
\end{equation}
Here I have suppressed the vector index because we are working in two dimensions, and thus $E^a$ is restricted to a one dimensional subspace, since it is a spacelike vector orthogonal to $\xi^a$.

Since $k$ and $\omega$ are functionally related, we can change variables to rewrite Eq. \eqref{genSol2} as,\footnote{Actually, since $\omega$ is proportional to the absolute value of $k$, there is a term missing in Eq. \eqref{genSolOm}, corresponding to the contributions to the integral in Eq. \eqref{genSol2} when $k<0$, i.e., for waves moving in the $-x$ direction.  We intentionally suppress that second term here, focusing only on waves moving in the positive $x$ direction.  Of course, an identical argument can be run for waves moving in the opposite direction.}
\begin{equation}\label{genSolOm}
E(x,t) = \frac{1}{\sqrt{2\pi}}\int_{\mathbb{R}} d \omega \tilde{C}(\omega) e^{i(k(\omega)x-\omega t)}=\frac{1}{\sqrt{2\pi}}\int_{\mathbb{R}} d \omega \tilde{C}(\omega) e^{i(\frac{\omega n(\omega)}{c}x-\omega t)}.
\end{equation}
Here we have absorbed the terms arising from the change of variables into $\tilde{C}(\omega)$, which can then be defined directly as an inverse Fourier transform:
\begin{equation}\label{Cdef}
\tilde{C}(\omega)=\frac{1}{\sqrt{2\pi}}\int_{\mathbb{R}} d t E(0,t)e^{i\omega t}.
\end{equation}
We now suppose that, whatever else may be the case, for all times $t< 0$, $E(0,t)=0$.  This assumption is meant to capture the idea that we are modeling a wave moving in the $+x$ direction that reaches the $x=0$ plane \emph{no earlier than} $t=0$.  Note that we are explicitly insisting that ``arrival,'' here, corresponds to a non-vanishing value of $E(x,t)$.  As we have seen, Sommerfeld thought of this ``arrival'' in terms of a step-function, so that one might take $E(x,0)>0$ for all $x<0$.  But for present purposes, the spatial form of $E(x,t)$ does not matter, so long as the assumption above holds (and so long as we restrict to waves moving in the $+x$ direction).  Indeed, one could allow the field to be smooth.


Sommerfeld's argument can now be cast in terms of properties of the integral on the right-hand side of Eq. \eqref{genSolOm}. The argument require four assumptions.  The first two concern the index of refraction $n(\omega)$.
\begin{enumerate}[\bf {Assumption} 1:] \singlespacing
\item The index of refraction $n(\omega)$, extended to a function of a complex variable (with the same functional expression), is analytic in the upper half of the complex $\omega$ plane.\label{analyticn}
\item The index of refraction $n(\omega)$ approaches $1$ uniformly as $|\omega|$ approaches $\infty$, for $\omega$ in the upper half of the complex $\omega$ plane.\label{biggun}
    \newcounter{enumTemp}
    \setcounter{enumTemp}{\theenumi}
\end{enumerate}
The remaining two assumptions regard the spectral amplitude, $\tilde{C}(\omega)$.
\begin{enumerate}[\bf {Assumption} 1:]
\setcounter{enumi}{\theenumTemp} \singlespacing
\item The spectral amplitude $\tilde{C}(\omega)$, extended to a function of a complex variable (with the same functional expression), is analytic in the upper half of the complex $\omega$ plane.\label{analyticc}
\item The spectral amplitude $\tilde{C}(\omega)$ approaches $0$ uniformly as $|\omega|$ approaches $\infty$, for $\omega$ in the upper half of the complex $\omega$ plane.\label{Cto0}
\end{enumerate}
The argument then relies on the following (slight strengthening) of Jordan's lemma, a classical result from complex analysis.\footnote{Compare, for instance, with \citet[p. 272-4]{Brown+Churchill}.  The present strengthened version of the result is proved in \ref{Proof}.}
\begin{prop}[Jordan]\label{Jordan} Given a complex-valued continuous function $f(z)$, if (1) $f(z) = g(z) e^{ia(z)z} $ for all $z\in C_R=\{z : z = Re^{i\theta},\theta\in [0,\pi]\}$, where (2) $a(z)$ is such that there exists some $\epsilon>0$ such that $a(z)>\epsilon$ for all $z\in C_R$ for sufficiently large $R$ and (3) $g(z)$ approaches $0$ uniformly as $|z|$ approaches $\infty$ in the upper half of the complex plane, then
\begin{equation}
\lim_{R\rightarrow\infty}\int_{C_R} f(z)dz=0.
\end{equation}
\end{prop}

We apply the result to the current case as follows.  We first define $a(\omega,x,t)=n(\omega)x/c-t$ and $g(\omega)=\tilde{C}(\omega)/\sqrt{2\pi}$, so that the integrand in Eq. \eqref{genSolOm} may be written as
 \[
I(\omega,x,t) = g(\omega) e^{ia(\omega,x,t)\omega}.
\]
Now consider (fixed) $x$ and $t$ such that $x/c-t>0$.  By assumption \ref{biggun}, $n(\omega)$ approaches $1$ uniformly as $|\omega|\rightarrow\infty$ in the upper half of the complex $\omega$ plane, and thus, given any $\delta > 0$, we can always find some $\Omega$ such that whenever $|\omega|>\Omega$ in the upper half of the complex plane, $1+\delta > n(\omega) > 1-\delta$.  Suppose, without loss of generality, that $x\geq 0$.\footnote{If $x\leq 0$, then an analogous argument with the same conclusion holds, beginning with the observation that for sufficiently large $|\omega|$, $a(\omega,x,t) \geq (1+\delta)x/c -t$.}  Then, for sufficiently large $|\omega|$, $a(\omega,x,t)\geq(1-\delta)x/c-t = x/c-t - \delta x/c$.  But since $x/c-t > 0$, we can always choose $\delta$ small enough that $x/c -t > \delta x /c$.  Thus there is an $\epsilon =(x/c-t - \delta x/c)/2 >0$ such that $a(\omega,x,t)>\epsilon$ for all sufficiently large $|\omega|$ in the upper half of the complex plane.  Invoking assumption \ref{Cto0} and the Jordan's lemma, we may conclude that whenever $x/c-t>0$,
\begin{equation}\label{Ito0}
\lim_{R\rightarrow\infty}\int_{C_R}I(\omega,x,t)d\omega=0,
\end{equation}
where, as in the proposition, $C_R$ is a semi-circular contour of radius $R$ in the upper half of the complex plane.

Eq. \eqref{Ito0} implies that whenever $x/c-t>0$, adding $\lim_{R\rightarrow\infty}\int_{C_R}I(\omega,x,t)d\omega$ to $E(x,t)$ contributes nothing. Thus,
\begin{align}
E(x,t) &= E(x,t) + \lim_{R\rightarrow\infty}\int_{C_R}I(\omega,x,t)d\omega \\
&=\frac{1}{\sqrt{2\pi}} \lim_{R\rightarrow\infty} \left(\int_{-R}^{R} d \omega \tilde{C}(\omega) e^{i(\frac{\omega n(\omega)}{c}x-\omega t)} + \int_{C_R} d \omega \tilde{C}(\omega) e^{i(\frac{\omega n(\omega)}{c}x-\omega t)}\right)\\
&= \frac{1}{\sqrt{2\pi}} \lim_{R\rightarrow\infty} \oint_{\gamma_R}\tilde{C}(\omega) e^{i(\frac{\omega n(\omega)}{c}x-\omega t)},
\end{align}
where $\gamma_R$ is the closed contour constructed by appending $C_R$ to the interval $[-R,R]$ on the real line.  But now note that, by assumptions \ref{analyticn} and \ref{analyticc}, $\tilde{C}(\omega) e^{i(\frac{\omega n(\omega)}{c}x-\omega t)}$ is everywhere analytic within the region enclosed by $\gamma_R$, for any $R>0$.  Thus, by appeal to Cauchy's integral theorem, we know that for every $R>0$,
\begin{equation}
\oint_{\gamma_R}\tilde{C}(\omega) e^{i(\frac{\omega n(\omega)}{c}x-\omega t)} = 0.
\end{equation}
So we can conclude that
\begin{equation}
E(x,t) = 0
\end{equation}
whenever $x > ct$---i.e., $E(x,t)$ vanishes everywhere to the right of the null line $x=ct$.  It is this result that is taken to rule out superluminal propagation of an electromagnetic field in a dielectric medium, since it shows that if an electromagnetic field is propagating in the $+x$ direction, and it vanishes for $x>0$ and $t=0$, then it vanishes for all $x>ct$.

We are now in a position to say why Sommerfeld's no-go result does not conflict with the calculation in section \ref{Gordon}.  The reason is simple: the two arguments make incompatible assumptions.  Specifically, the argument in section \ref{Gordon} assumes that the index of refraction $n$ is independent of frequency, and it finds superluminal propagation only when $n<1$ for all $\omega$.  Sommerfeld's argument, meanwhile, allows $n$ to vary with $\omega$---but it requires that for sufficiently large frequency, $n(\omega)$ approach $1$.  This explicit assumption of Sommerfeld's argument---assumption \ref{biggun}---rules out the case studied in section \ref{Gordon} and relieves any apparent tension.

Indeed, one can say a bit more.  First of all, we saw in section \ref{Gordon} that in the case where $n\geq 1$, the causal cones associated with the electromagnetic field in the medium were always no wider than the metric lightcones, and thus the argument above agrees with Sommerfeld's argument in the case where both apply.  Moreover, if one weakens assumption \ref{biggun} of Sommerfeld's argument slightly, and insists only there exists some constant $N$ such that $n$ approach $N$ uniformly as $|\omega$ approaches $\infty$ in the upper half of the complex plane, without requiring that $N=1$, then Sommerfeld's argument would still go through---except that the result would be that $E(x,t)=0$ whenever $x > ct/N$.  And in particular, if $N<1$ , Sommerfeld's argument does not rule out propagation of the field at the superluminal velocity $c/n$.  And so once again, we see agreement between the argument of section \ref{Gordon} and Sommerfeld's argument in a (different) case where the assumptions of both apply.

It seems, then, that what we make of Sommerfeld's no-go result turns on the status of the assumptions underlying it---particularly assumption \ref{biggun}.\footnote{The other three are more mild.  Assumptions \ref{analyticc} and \ref{Cto0}, for instance, may be derived by requiring that $E(0,t)$ and its first derivative are bounded \citep[\S 7.1]{Oughstun+Sherman}.  Assumption \ref{analyticn}, meanwhile, may be derived, in the presence of some other modest assumptions, from the requirement that a medium cannot ``respond'' to an electromagnetic field prior to the arrival of the field.  (See \citet[\S 7.10]{Jackson} for an extended discussion.)  This latter assumption is also sometimes thought of as a kind of causality requirement---an effect may not precede its cause---albeit one of a different character than the ``relativistic causality'' requirements under consideration in the present paper.  That said, there is a sense in which the ultimate justification for assumption \ref{analyticn}---and the justification that \citet[p. 337]{Jackson} ultimately relies on---is the same as the justification for assumption \ref{biggun}: both assumptions are met by standard ``realistic'' models of matter.  Thus, though I do not pursue this line here, one might also put pressure on whether assumption \ref{analyticn} has anything to do with ``relativistic'' considerations.}  The assumption is usually justified by physical considerations.  For instance, one might argue that atoms have characteristic length scales associated with them; arbitrarily high frequency waves have short wavelengths (assuming the index of refraction is well behaved), and waves with wavelengths much shorter than the characteristic length of an atom will not interact strongly, so that the waves do not ``see'' the atoms \citep[p. 13]{Milonni}.  Similarly, one might argue that atoms interact with light only near atomic resonance frequencies; sufficiently high-frequency light ``misses'' these frequencies \citep[pp. 313-4]{Jackson}.  Or one might reason that high frequency waves are changing so rapidly that the matter does not have time to react, so there is no interaction \citep[p. 267]{Landau+Lifshitz}. The upshot of all of these arguments is that matter should become transparent to light of sufficiently high frequency, and so $n(\omega)$ should approach $1$ as $|\omega|$ approaches $\infty$.

Of course, such arguments are heuristic, though they appear to be supported by standard modeling methods \citep[\S 7.5]{Jackson}.  But I will not discuss them further.  For present purposes, one might just stipulate that such arguments do justify assumption \ref{biggun}, and even that they explain why we do not appear to observe superluminal propagation of electromagnetic fields in real media.  The important point is that, whatever else might be the case, such arguments have \emph{nothing} to do with relativity theory.  In other words, there is no sense in which the crucial assumption of Sommerfeld's theorem follows from, or is otherwise imposed by, relativistic considerations.  It is justified by appeal to the atomic theory of matter and models of interactions between electromagnetic fields and atomic resonances.  Even if this justification is sound and convincing, it is some collection of assumptions about the nature of matter that do the work in ruling out superluminal wavefront velocities on Sommerfeld's argument.  And as we have seen, if we relax this assumption, even Sommerfeld's argument appears to allow for superluminal propagation.

\section{Reservations and Prospectives}\label{woah}

I have now made the principal arguments of the paper.  I have defended three theses: (1) the Geroch-Earman criterion of superluminal propagation may be understood to make precise, and to generalize, the sense of superluminal propagation given by \citet{Sommerfeld}; (2) by the Geroch-Earman criterion of superluminal propagation (and thus, in a straightforward sense, by the Sommerfeld criterion), for some parameter values, the macroscopic Maxwell equations exhibit superluminal propagation; and (3) the widely cited \citet{Sommerfeld} no-go result crucially depends on an assumption that bears no apparent connection to relativity theory, and indeed, relaxing that assumption appears to allow superluminal propagation.

None of these theses imply that any real physical systems exhibit superluminal propagation under realistic conditions; indeed, it would seem real physical systems satisfy the assumptions of Sommerfeld's theorem.  But I believe the arguments given here \emph{do} support the more modest claim that our understanding of the relationship between relativity theory and superluminal propagation requires further study---and in particular, that insofar as we have convincing arguments that electromagnetic fields do not propagate superluminally in a medium, relativity theory plays no role.

Still, even this modest moral should be hedged.  There are various well-known senses in which superluminal propagation in relativity theory may seem pathological.  For instance, some timelike observers will say that a superluminal field is propagating instantaneously or even backwards in time, since the surfaces of constant phase are spacelike.  Similarly it is not clear that one can coherently associate properties such as mass or 4-momentum to such fields.  One might take such considerations to reflect, or even amount to, a general sort of incompatibility between relativity and superluminal propagation, of a sort that falls short of outright contradiction---and for that reason, perhaps cannot be captured in a clean no-go theorem---but which nevertheless leads to significant theoretical problems. In other words, even if \citet{Geroch2} is correct that there is a sense in which superluminal propagation is compatible with the geometry of Minkowski spacetime, one might worry that our standard interpretation of the physical significance of that geometry is so severely undermined by superluminal propagation as to render relativity theory unusable or otherwise unacceptable.

If one adopts such a line, then Sommerfeld's argument might be taken to have a different status. It shows that, taken as a whole, our theory of electromagnetism and its interactions with matter, given physically reasonable assumptions, does not lead to the kinds of incoherence described above.  From this perspective, one should not care that relativity theory plays no obvious role in Sommerfeld's argument.  Relativity theory rules out superluminal propagation insofar as, given the pathologies just mentioned, we would find any theory that permitted superluminal propagation theoretically unsatisfactory.  So Sommerfeld shows that, as a matter internal to electromagnetism, our theory of the interactions between electromagnetic fields and matter is not unsatisfactory in this particular way.

Still, if the pathologies connected with superluminal propagation fall short of outright contradiction with relativity, it would be desirable to identify more clearly what the problems are and why we find them troubling.  (As \citet{Geroch2} points out, the world might well work in ways that we find troubling!)  One place where we might expect problems to arise would concern the way superluminal fields interact with other matter; another place might be in how such fields behave in general relativity, where they would contribute as sources in Einstein's equation.  There is one straightforward sense in which one can treat interactions between superluminal fields and other fields: namely, by introducing appropriate interaction terms in the relevant differential equations.  But if one wants to provide a full treatment of interactions between fields that makes contact with the rest of physics, one needs to understand the energy-momentum content of superluminal fields, and one needs to understand how energy-momentum is transferred between fields.  Similarly, one needs to be able to associate an energy-momentum tensor with a superluminal field in order for that field to be a source term in Einstein's equation.

Energy-momentum considerations are especially salient in the present case.  For instance, it would be valuable to know whether the electromagnetic fields described in section \ref{Gordon} violate the standard energy conditions, which are often taken as a litmus test for superluminal propagation of energy-momentum.\footnote{For more on energy conditions, see \citet[\S 2.5]{MalamentGR} or \citet{CurielEC}.}  (One suspects they do, but it is worth studying.) I have said nothing to address this issue.  The reason is that there is a century-old, still-unresolved puzzle related to the energy-momentum content of an electromagnetic field in a moving dielectric.\footnote{See \citet{Abraham1,Abraham2}, \citet{Minkowski}, and \citet{Pfeifer+etal}.}  Specifically, several proposals have been made for what the energy-momentum tensor of an electromagnetic field in a moving dielectric should be, and there is no consensus in the physics literature concerning which proposal is correct.  Describing, never mind adjudicating, this debate would take the present paper too far afield, and so I have postponed any discussion of energy-momentum to future work.

That said, I will mention one reason that the outcome of such a study would be of interest.  Suppose that the energy-momentum associated with an electromagnetic field in a dielectric \emph{does} violate one of the standard energy conditions, such as the so-called dominant energy condition.\footnote{The dominant energy condition holds of an energy-momentum tensor $T^{ab}$ just in case either $T^{ab}=\mathbf{0}$ or else for any future-directed timelike vector $\eta^a$ at any point $p$, $T^{ab}\eta_b$ is future-directed and causal.}  Then it would seem there is a problem.  One can show that the energy-momentum tensor associated with solutions of the vacuum Maxwell equations always \emph{does} satisfy the dominant energy condition.  Moreover, one can show that if two energy-momentum tensors both satisfy the dominant energy condition, then so does their sum---and so, conversely, if the total energy-momentum tensor fails to satisfy the dominant energy condition, at least one of the contributing energy-momentum tensors must fail to satisfy it.\footnote{I am grateful to David Malament for suggesting this point.}  Thus, if interactions between electromagnetic fields and dielectric media are treated in the same way that other interactions in relativity theory are, that is, by adding energy-momentum tensors, it would seem to follow that interactions with a medium could produce violations of the dominant energy condition only if the medium itself violates the dominant energy condition.  In other words, one might think one gets a certain kind of superluminality in electromagnetic fields only for media that \emph{already} exhibit a kind of superluminality.

There are several places where this last argument could go wrong.  In particular, \citet{Earman} has shown that superluminal propagation in the Geroch-Earman sense and violations of the dominant energy condition are not as tightly linked as one might have expected.  Moreover, it is not clear that interactions between electromagnetic fields and matter of the form captured by the macroscopic Maxwell equations are naturally represented by summing two energy-momentum tensors.  But such considerations only provide more reasons that a full analysis of energy-momentum is necessary before the examples described in the present paper are fully understood.

I will conclude by mentioning four other worries I have about the argument in the present paper.  The first concerns the foundations of the macroscopic Maxwell equations in the first place.  Eqs. \ref{MacroMaxwell2} are textbook equations of motion, but they are not taken to be ``fundamental'', and their derivation requires a number of assumptions.  Although there is a sense in which they are ``relativistic'' equations, one might still wonder whether the arguments by which they are derived are compatible with the spirit of relativity theory.  If not, then the example given in section \ref{Gordon} would be of largely formal interest: there would be certain phenomenological equations with undesirable properties in unphysical regimes, which would be neither surprising nor all that exciting.  This suggests that some sustained philosophical attention should be paid to the textbook derivations of these equations.

The second worry is related.  In the course of the analysis given in section \ref{Gordon}, I assumed that the 4-velocity of the dielectric, $\xi^a$, was constant.  But as I pointed out in fn. \ref{momentum}, insofar as the medium affects the behavior of the electromagnetic fields within the medium, conservation considerations would lead one to expect that the medium will be affected by the fields, perhaps through a change in the 4-velocity of the medium.  The macroscopic Maxwell equations do not capture this feature of the interaction, which again might be taken to undermine the significance of any arguments based on them.  Here, too, a detailed treatment of the energy-momentum properties of the fields and the medium would be helpful.

The third worry is that although the arguments surveyed at the end of section \ref{Sommerfeld} for why $n$ approaches $1$ for large frequency do not appear to be motivated by relativistic considerations, it may be that there is some sense in which relativity theory does provide constraints on the index of refraction for high-frequency waves.  I do not see how such an argument might go, but it is presumably within the realm of logical possibility.\footnote{Erik Curiel (private correspondence) has suggested that one might think of higher frequency waves as carrying more momentum; thus, it would require unbounded work for a medium to deflect arbitrarily high frequency waves.  This may be a fruitful line to pursue, though I worry that it depends crucially on an intuition from quantum mechanics, concerning the relationship between energy and frequency.  Classically, the energy-momentum of an electromagnetic field depends on the field strength, i.e., the amplitude of the wave, and so could be made arbitrarily small, relative to any given observer, for arbitrarily high frequencies.}

Finally, the fourth worry, which I briefly mentioned in footnote \ref{worries}, is that Einstein's equation plays no apparent role in the Geroch-Earman condition or in the calculations in section \ref{Gordon}.  Of course, the behavior of ``test fields,'' i.e., fields that do not act as sources in Einstein's equation, is of some independent interest.  But if one is to claim that superluminal propagation is in some sense compatible with relativity, one should also ask whether the sorts of interactions between matter and geometry that are captured by Einstein's equation put constraints on the possible superluminal propagation of matter.  One way to think about this issue within the framework set by the Geroch-Earman condition would be to consider a system of coupled hyperbolic differential equations that includes Einstein's equation and then find the causal cones associated with the entire system.  I do not know how this changes things, in general or in the special case of the macroscopic Maxwell equations, but it seems to me that addressing this question would be a crucial next step in understanding the basic issues raised in this paper.

In a sense, however, all of these reservations are grist for my mill, insofar as the ultimate claim is that the relationship between relativity theory and superluminal propagation is not well understood.  Progress on any of the routes I have suggested in this section would certainly contribute to that understanding.

\appendix

\section{Proof of Prop. \ref{Jordan}}\label{Proof}

The version of Jordan's lemma used in section \ref{Sommerfeld} is stronger than one usually encounters in complex analysis textbooks.  For completeness, and because this strengthened version is invoked without proof or reference, in, for instance, \citet[\S 7.11]{Jackson}, I present a proof here.
\setcounter{thm}{2}
\begin{prop}[Jordan]\label{Jordan} Given a complex-valued continuous function $f(z)$, if (1) $f(z) = g(z)e^{ia(z)z}$ for all $z\in C_R=\{z : z = Re^{i\theta},\theta\in [0,\pi]\}$, where (2) $a(z)$ is such that there exists some $\epsilon>0$ such that $a(z)>\epsilon$ for all $z\in C_R$ for sufficiently large $R$ and (3) $g(z)$ approaches $0$ uniformly as $|z|$ approaches $\infty$ in the upper half of the complex plane, then
\begin{equation}
\lim_{R\rightarrow\infty}\int_{C_R} f(z)dz=0.
\end{equation}
\end{prop}
Proof. Let $f(z)$ be as described in the proposition.  It follow that
\[
\int_{C_R}f(z)dz=iR\int_0^{\pi}g(Re^{-\theta})e^{iRa(Re^{i\theta})(\cos\theta+i\sin\theta)}e^{i\theta}d\theta.\]
Then, using the fact that $|\int_{a}^bf(x)dx|\leq\int_a^b |f(x)|dx$, we have that
\[
\left|\int_{C_R}f(z)dz\right|\leq R\int_0^{\pi}|g(Re^{i\theta})|e^{-Ra(Re^{i\theta})\sin\theta}d\theta\leq Rg_R\int_0^{\pi}e^{-Ra(Re^{i\theta})\sin\theta}d\theta\]
where $g_R=\max_{\theta\in [0,\pi]} |g(Re^{i\theta})|$.  Now observe that $e^{R a(Re^{i\theta})\sin\theta}\leq e^{-R a_R\sin\theta}$, where $a_R=\min_{\theta\in [0,\pi]}a(Re^{i\theta})$, and so
\[
\left|\int_{C_R}f(z)dz\right|\leq Rg_R\int_0^{\pi}e^{-Ra_R\sin\theta}d\theta.
\]
Now, invoking Jordan's inequality \citep[p. 273]{Brown+Churchill}, which states that for $R>0$,
\[
\int_0^{\pi}e^{-R\sin\theta}d\theta < \frac{\pi}{R},
\]
and the fact that for all sufficiently large $R$, $a_R>0$, we find that for sufficiently large $R$,
\[
\left|\int_{C_R}f(z)dz\right|\leq \pi\left(\frac{g_R}{a_R}\right).
\]
Finally, using the facts that for large $R$, $a_R$ is bounded away from $0$, and that $\lim_{R\rightarrow \infty} |g_R|=0$, we conclude that
\[
\lim_{R\rightarrow\infty}\left|\int_{C_R}f(z)dz\right|=0.\]
The proposition follows immediately.\hspace{.25in}$\square$

\section*{Acknowledgments}
This material is based upon work supported by the National Science Foundation under Grant No. 1331126.  I am grateful to Jeff Barrett, Sam Fletcher, Bob Geroch, David Malament, John Norton, Steven Sagona-Stophel, Chris Search, Chris Smeenk, and Sheldon Smith for helpful conversations on topics related to the ones discussed here.  I am also grateful to helpful audiences at the UC Irvine Joint Particle Seminar and the University of Western Ontario's 2013 Philosophy of Physics Conference for comments and discussion.  Jeremy Butterfield, Erik Curiel, David Malament, John Manchak, Ben Feintzeig, Sam Fletcher, Marian Rogers, and an anonymous referee deserve special thanks for helpful---and in the cases of Curiel and Butterfield, exhaustive---comments on earlier drafts of the paper.

\singlespacing
\bibliography{superluminal}
\bibliographystyle{elsarticle-harv}

\end{document}